\documentclass[preprint2]{emulateapj}

\usepackage{bm}
\usepackage{times}
\usepackage{natbib}
\newcommand{\EQ}{\begin{equation}}
\newcommand{\EN}{\end{equation}}

\newcommand{\uu}{\mbox{\boldmath $u$} {}}

\newcommand{\JJ}{\mbox{\boldmath $J$} {}}
\newcommand{\BB}{\mbox{\boldmath $B$} {}}
\newcommand{\AAA}{\mbox{\boldmath $A$} {}}

\newcommand{\SSSS}{\mbox{\boldmath ${\sf S}$} {}}
\newcommand{\nab}{\mbox{\boldmath $\nabla$} {}}

\newcommand{\ve}[1]{\boldsymbol{#1}}%
\newcommand{\kef}{k_{\rm f}}
\newcommand{\DD}{{\rm D} {}}

\newcommand{\K}{\,{\rm K}}
\newcommand{\g}{\,{\rm g}}
\newcommand{\s}{\,{\rm s}}
\newcommand{\cm}{\,{\rm cm}}
\newcommand{\km}{\,{\rm km}}
\newcommand{\kms}{\,{\rm km/s}}
\newcommand{\pc}{\,{\rm pc}}
\newcommand{\kpc}{\,{\rm kpc}}

\newcommand{\Gyr}{\,{\rm Gyr}}
\newcommand{\erg}{\,{\rm erg}}
\newcommand{\dyn}{\,{\rm dyn}}
\begin{document}
\title{Dynamo action in thermally unstable interstellar flows}
\author{Maarit J.\ Mantere \& Elizabeth Cole}
\altaffiltext{}{Department of Physics, PO
  BOX 64, FIN-00014 University of Helsinki, Finland}

\begin{abstract} Numerous studies have investigated the role of
  thermal instability in regulating the phase transition between the
  cold cloudy and warm diffuse medium of the interstellar
  medium. Considerable interest has also been devoted to investigating
  the properties of turbulence in thermally unstable flows, with a
  special emphasis on molecular clouds and the possibility of star
  formation. In this study, we investigate another setting in which
  this instability may be important, namely its effect on dynamo
  action in interstellar flows. The setup we consider is a three
  dimensional periodic cube of gas with an initially weak magnetic
  field, subject to heating and cooling, the properties of which are
  such that thermal instability is provoked in a certain temperature
  regime. Dynamo action is established through external forcing on the
  flow field. By comparing the results with a cooling function with
  exactly the same net effect but no thermally unstable regime, we
  find the following. Reference runs with non-helical forcing were
  observed to produce no small-scale dynamo action below the Reynolds
  number 97. Therefore, we expect the magnetic fields generated in the
  helical runs to be purely due to the action of a large-scale dynamo
  mechanism. The critical Reynolds number for the onset of the
  large-scale dynamo was observed to roughly double between the
  thermally stable versus unstable runs, the conclusion being that the
  thermal instability makes large-scale dynamo action more
  difficult. Whereas density and magnetic fields were observed to be
  almost completely uncorrelated in the thermally stable cases
  investigated, the action of thermal instability was observed to
  produce a positive correlation of the form $B \propto
  \rho^{0.2}$. This correlation is rather weak, and in addition it was
  observed to break down at the limit of highest densities.
\end{abstract}

\section{Introduction}

The role of thermal instability (hereafter TI) in regulating the phase
transition between the cold and warm neutral components, normally
denoted with CNM and WNM, of the interstellar matter (hereafter ISM)
has been intensively studied over several decades \citep[for a review,
see e.g.][]{Coxreview05} since the original seminal paper by
\citet*{FGH69} proposed this mechanism, known as the
FGH-model. According to it, two thermally stable phases (cold and
cloudy; warm and diffuse) co-exist in pressure equilibrium regulated
by the presence of a thermally unstable phase at an intermediate
temperature. Now it is observationally well-established that the
picture is not quite simple as that, but there is also a third, hot
and dilute, component of the ISM. The first model that incorporated
the hot component, generated by the action of supernova explosions
(hereafter SNe), to the FGH model was presented by \citet*{MO77}. This
model suggested rather drastic modifications to the structure of ISM:
with the inclusion of SN-action, most (70--80 percent) of the gas
became filled by the hot component. Since then, with extensive
numerical modelling of the SN-driven flows
\citep[e.g.][]{CP85,Rosen93,Korpi99,GP99,Avillez00,Balsara04,Gressel08}
and more careful observational determinations \citep[for a review, see
e.g.][]{Ferriere01}, the estimates of the filling factor of the hot
component have been reduced to 10--30 percent near the Galactic
midplane, being bigger at larger heights. Moreover, most of the hot
gas seems to be confined in large bubbles created by clustered SN
activity rather than being distributed homogeneously around the Galaxy
\citep[e.g.][]{Ferriere01}.

A number of investigations of interstellar turbulence at relevant
scales for studying molecular cloud formation or any other process of
interest in the atomic ISM still use the FGH-picture as a starting
point. This seems justified in the light of the hot gas filling factor
being relatively small near the Galactic midplane. Various numerical
studies investigating the interaction of turbulence and TI have been
published. In most of these studies turbulence is forced by sources
other than the TI itself: random turbulent forcing at varying scales
and Mach numbers \citep[e.g.][]{Gazol05}, localized injections of
energy mimicking stellar winds \citep[e.g.][]{VS00}, the
magnetorotational instability \citep*[e.g.][]{Piontek07a}, and
systematic large-scale motions such as propagating shock fronts
\citep*[e.g.][]{KI02} and converging flows e.g. by \citet*{AH05},
\citet{Heitsch05} and \citet{VS06} have been considered. One of
the major findings from these models is that, due to the turbulence
present in the system, large pressure deviations are generated and
significant amounts of gas can exist in the thermally unstable
regime. These results suggest, as already noted by \citet*{NF96}, that
the FGH picture of the ISM exhibiting 'discrete' temperatures and
densities and a unique equilibrium pressure should be modified in the
direction of a 'continuum' of states with an overall pressure balance
but with large deviations from it.

In recent years the possibility of driving turbulence by the TI itself
has also received some attention. Contrary to \citet*{KN02}, who found
turbulence to die out as a power law, \citet*{IK07} found the
turbulence to be sustained---at least for times up to $0.1\Gyr$,
although the turbulent motions were too weak to be significant in the
ISM. However, in the study by \citet*{BKM07}, hereafter BKM07, the
TI-generated turbulence was observed to decay in simulations extending
to more than $1\Gyr$. Even the inclusion of large-scale shear did not
help sustain the turbulence. Therefore, it seems evident that TI alone
cannot produce self-sustained turbulence, in contrast to the
gravitational instability studied by \citet{Gammie01} or the
magnetorotational turbulence studied e.g. by \citet*{HGB95} and
\citet{BNST95}.

The line-of-sight magnetic field strengths in dense molecular clouds,
as revealed by the Zeeman splitting measurements of molecular lines,
follow an approximate correlation of the form $B_{||} \propto
n(H_2)^{1/2}$, where $n(H_2)$ is the density of the molecular hydrogen
in the cloud, e.g. \citet{Crutcher99} \citep*[for a recent review, see
e.g.][]{Heiles_et_Crutcher05}. For diffuse clouds at lower densities n
$\sim$ 0.1 -- 100 cm$^{−3}$ also seen through the Zeeman splitting of the
HI line, no clear correlation between the magnetic field strength and
density has been found
\citep*{Troland_et_Heiles86,Myers95,Heiles_et_Crutcher05}.
\citet*{PVS03} suggested that this phenomenon could be due to the
strongly varying Alfv\'enic Mach number, $\rm Ma_a$, in the turbulent
ISM. At low values of $\rm Ma_a$, their analysis revealed an
anti-correlation between the density and magnetic pressure due to the
dominance of slow MHD waves. At higher $\rm Ma_a$, both slow and fast
modes were possible, their interference leading to de-correlation for
intermediate densities, and re-correlation for the highest
densities. \citet{Heitsch04} proposed an alternative mechanism, namely
ambipolar diffusion, enhanced by the vigorous interstellar turbulence
to be rapid enough to de-correlate the magnetic field from the
density. The fast reconnection of magnetic field lines in the cloudy
medium has also been discussed in this context, e.g. by \citet{SL10}.
 
In addition, there is observational evidence of it being particularly
the {\it regular} component, seen through synchrotron emission, that
is de-correlated from the gas density. For instance, observations of
barred spiral galaxies indicate the magnetic fields forming coherent
structures over the length scale of the bars themselves are not
reacting as strongly as expected to the compression by the systematic
motions occurring in the bars (e.g. NGC 1097 and NGC 1365 analyzed by
Beck et al., 2005). A similar situation has been reported in M51 by
\citet{Fletcher10}, where the regular magnetic field is not strongest
at the locations of the highest compression due to the spiral density
wave. It has been proposed that such a loose or completely nonexistent
correlation between the regular component of the magnetic field and
the diffuse clouds could be a result of the detachment of the
large-scale field from the clouds in a fragmentation process such as
the one resulting from thermal or gravitational instability
\citep{Beck05}.

During the recent years, numerical models in the magnetohydrodynamic
regime including thermal and gravitational instabilities able to reach
down to the size of the molecular clouds themselves have been
developed by e.g. \citet{Hennebelle08}, \citet*{HH08} and
  \citet{Banerjee09}. The results from these calculations seem to be
in very good agreement with the observations, i.e. producing very weak
correlation between the magnetic field and density for lower
densities, while re-establishing the correlation of the form $B
  \propto n^{1/2}$ at higher densities. Our first motivation in this
manuscript is to isolate the role of thermal instability in the
process that produces the density-magnetic field correlation.

Another interesting question that has not gained much attention is
the effect of thermal instability on the galactic dynamo
mechanism. Evidently, thermal instability cannot be expected by itself
to drive turbulence that could generate and maintain a self-sustained
dynamo. Rather, it can be expected to influence the dynamics of a flow
that is dynamo active, and therefore may also have an effect on the
properties of the dynamo. This is the second aspect we are aiming at
investigating in detail in the present manuscript. 

We adopt a very simple approach: as in our previous study of thermal
instability under a purely hydrodynamic setting (BKM07), we use a
cubic periodic computational domain filled with an initially uniform,
non-stratified gas that has the characteristics of the atomic
component of the interstellar matter. Unlike BKM07, a weak random
magnetic field is inserted into the system at the initial state. Dynamo
action is achieved by using an external forcing function on the flow
field. In the case where the forcing is helical, a large-scale dynamo
develops. For non-helical forcings, only small-scale dynamo action can
be expected in our setting. The system is then subjected to either
thermally stable or unstable cooling functions balanced by a constant
UV-heating function. From the resulting set of models, we monitor the
growth rate and saturation properties of the different types of
dynamos, and the correlation between the density and magnetic field
developing in each system.

\section{Model}

\subsection{Governing equations}

We consider the governing equations for a compressible and magnetized perfect gas,
\EQ
{\DD\ln\rho\over\DD t}=-\nab\cdot\uu,
\EN
\EQ
\rho{\DD\uu\over\DD t}=-\nab p + \JJ\times\BB + \nab\cdot(2\nu\rho\SSSS)+ \bm{f}_{\rm force},
\label{dudt}
\EN
\EQ
T{\DD s\over\DD t}=2\nu\SSSS^2+{1\over\rho}\nab\cdot\left(c_p\rho\chi\nab T\right)-{\cal L},
\EN
\EQ
{\DD \AAA \over \DD t} = - (\nab \uu)^{\rm T}\AAA-\mu_0\eta \JJ, \label{dadt}
\EN
where $\uu$ is the velocity, $\rho$ the density, $s$ the specific entropy, $\AAA$ the vector potential of the magnetic field $\BB=\nab\times\AAA$, $\JJ=\mu_0^{-1} \nab\times\BB$ the current,
with ${\sf S}_{ij}=\frac{1}{2}(u_{i,j}+u_{j,i})
-\frac{1}{3} \delta_{ij}\nab\cdot\uu$ being the traceless rate of strain tensor,
$\nu$ is the kinematic viscosity, $\chi$ is the thermal diffusivity, $\eta$ the magnetic diffusivity, $\mu_0$ the permeability of free space, and ${\cal L}$ is the net cooling/heating, i.e.\ the difference between cooling
and heating functions, with
\EQ
{\cal L}=\rho\Lambda-\Gamma,
\EN
where $\Gamma=\mbox{const}$ is assumed for the heating function; here
we consider the photoelectric heating by interstellar grains caused by
the stellar UV radiation field, for which Wolfire et al.\ (1995) give
the value of $0.015\erg\g^{-1}\s^{-1}$ at $n=1\cm^{-3}$.

Following common practice, we adopt a perfect gas where $\rho$ and
$s$ are related to pressure $p$ and temperature $T$ via the relations
\begin{equation}
p={{\cal R}\over\mu}\rho T,\quad
s=c_v\ln p-c_p\ln\rho+s_0,
\label{eos}
\end{equation}
where ${\cal R}=8.314\times10^7\cm^2\s^{-2}\K^{-1}$ is the universal
gas constant, and $\mu$ is the mean molecular weight; in this study we
have adopted $\mu=0.62$ in all cases, corresponding to complete
ionization. This is naturally unrealistic for the cold and cool
phases, for which the ionization fraction can be expected to
be low. To choose a higher mean molecular weight better corresponding
to the cooler phases, however, would be equally unrealistic, as a
  significant fraction of the gas is still expected to be ionized. A
better approach to deal with this issue would be to follow the
ionization fraction and adjust $\mu$ accordingly. This, however, is
out of the scope of the present study. The rest of the thermodynamic
quantities read ${\cal R}/\mu=c_p-c_v$, with $c_p$ and $c_v$ being the
specific heats at constant pressure and volume, respectively,
$\gamma=c_p/c_v=5/3$ is their assumed ratio, $c_{\rm s}$ is the
adiabatic sound speed, $T$ is the temperature, and the two are related
to the other quantities via $c_{\rm s}^2=\gamma{\cal R}T/\mu$.  The
specific entropy is defined up to a constant $s_0$, the value of which is
unimportant for the dynamics.

\begin{figure}[!t]
  \centering \includegraphics[width=0.98\columnwidth]{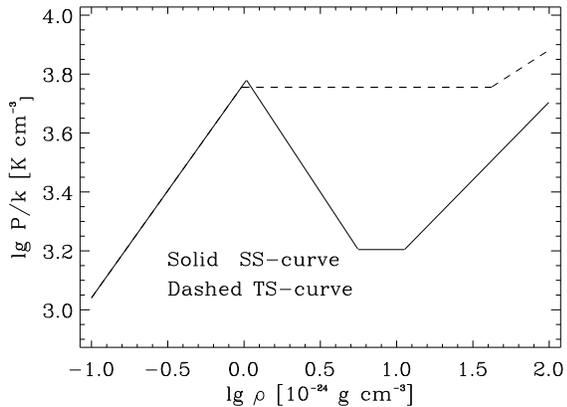}
  \caption{The equilibrium pressures calculated as function of density
    for the unstable 'SS' curve (solid) and stable 'TS' curve
    (dashed).}
  \label{fig:cooling}
\end{figure}

We adopt a parameterization of the cooling function equal to that
given by S{\'a}nchez-Salcedo et al.\ (2002), which has been obtained
by fitting a piecewise power law function of the form 
\EQ
\Lambda(T)=C_{i,i+1}T^{\beta_{i,i+1}} \quad\mbox{for $T_i\leq T<
  T_{i+1}$},
\label{CoolingCurve}
\EN to the equilibrium pressure curve of the standard model of Wolfire
et al.\ (1995) for the ISM in the solar neighborhood. This cooling
function has a thermally unstable branch in between the temperatures
$T=313$\,K and $6102$\,K, where $\beta < 1$; henceforth, we
denote this cooling function with 'SS'. To be able to extract the
effect of the thermally unstable branch of the cooling function on the
flow dynamics, we fabricate a thermally stable counterpart, denoted
with 'TS', with the criterion that for constant density, the net
cooling caused by these two functions is the same. The coefficients
for both cooling functions are given in Table~\ref{Tcooling}, and a
comparison plot is presented in Fig.~\ref{fig:cooling}. In this
figure, we plot the heating-cooling equilibrium pressures with the two
cooling functions, with respect to the chosen value of the UV-heating
function $\Gamma$.

We consider the case where turbulence is driven by an additional body
force, {\bf $\bm{f}_{\rm force}$}, in the momentum equation. Following
  \citet{KB09}, we use a forcing function $\bm{f}_{\rm force}$ given
  by
\begin{eqnarray} \label{ff}
\bm{f}_{\rm force}(\bm{x},t) = {\rm Re} \{N \bm{f}_{\bm{k}(t)} \exp [i \bm{k}(t)
  \cdot \bm{x} + i \phi(t) ] \}\;,
\end{eqnarray}
where $\bm{x}$ is the position vector, $N = f_0 c_{\rm s} (k c_{\rm
  s}/\delta t)^{1/2}$ is a normalization factor, $f_0$ is the forcing
amplitude, $k = |\bm{k}|$ the wavevector, $\delta t$ is the length of
the time step, and $-\pi < \phi(t) < \pi$ a random delta-correlated
phase. The wavevector and the random phase change at every time step,
resulting in a forcing function that is $\delta$-correlated in time. 
We force the system with transverse helical waves,
\begin{eqnarray}
\bm{f}_{\ve{k}} = {\sc R} \cdot \bm{f}_{\ve{k}}^{\rm nohel}\ \ {\rm with} \ \ {\sc R}_{ij} = \frac{\delta_{ij} - i \sigma \epsilon_{ijk} \hat{k}_k}{\sqrt{1 + \sigma^2}}.
\end{eqnarray}
The helical case with positive helicity corresponds to $\sigma=1$,
with the forcing function
\begin{equation}
\bm{f}_{\ve{k}}^{\rm nohel} = \left( \bm{k} \times \hat{\bm{e}}\right)/\sqrt{\bm{k}^2 - \left( \bm{k} \cdot \hat{\bm{e}} \right)^2},
\end{equation}
where $\hat{\bm{e}}$ is an arbitrary unit vector not aligned with the
forcing wavevector. 

Thus, $\bm{f}_{\bm{k}}$ can be chosen to give either non-helical
($\sigma=0$) or helical ($\sigma=1$) transversal waves with
$|\bm{f}_{\ve{k}}|^2 = 1$, where ${\bm k}$ is chosen randomly from a
predefined range in the vicinity of the average non-dimensional
wavenumber $\kef/k_1$ at each time step. Here $k_1$ is the wavenumber
corresponding to the domain size, and $\kef$ is the wavenumber of the
energy-carrying scale, chosen to be between 2.5 and 3.5 times the
smallest wavenumber in the box, $k_1=2\pi/(0.2\kpc)$. In a system very
similar to the one under study here, namely \citet{Axel01},
large-scale dynamo action was found with helical forcing. In the
presence of shear, non-helical forcing has also been found to excite a
large-scale dynamo \citep[e.g.][]{Yousef08}, but in the absence of it,
can lead only to small-scale dynamo action \citep[e.g.][]{HBM04}. In
this study, both types of forcings are investigated, in the hope to
study the effect of TI both on the small- and large-scale dynamos. We
systematically vary the type and vigor of the forcing, and subject the
gas either to thermally stable or unstable cooling function; the
parameters for the runs produced can be found from
Table~\ref{table:runs}.

\begin{table}[t!]\caption{Coefficients for the thermally unstable (SS)
    and stable (TS) cooling curves.}\vspace{12pt}\centerline{\begin{tabular}{crcrcr}
$i$ & $T_i$ & $C^{U}_{i,i+1}$ & $\beta^{U}_{i,i+1}$ & $C^{S}_{i,i+1}$ & $\beta^{S}_{i,i+1}$ \\
\hline
1 &    10  & $3.42\times10^{16}$ & 2.12   & $2.16\times10^{17}$ & 1.50 \\
2 &   141  & $9.10\times10^{18}$ & 1.00   & $2.56\times10^{18}$ & 1.00 \\
3 &   313  & $1.11\times10^{20}$ & 0.56   & $2.56\times10^{18}$ & 1.00 \\
4 &  6102  & $2.00\times10^{ 8}$ & 3.67   & $2.00\times10^{8}$  & 3.67 \\
5 & $10^5$ & $7.96\times10^{29}$ &$-0.65$ & $7.96\times10^{29}$ &$-0.65$ \\
\label{Tcooling}\end{tabular}}\end{table}

It is convenient to measure time in gigayears (Gyr), speed in km/s,
and density in units of $10^{-24}\g\cm^{-3}$. The resulting unit of pressure is
$10^{-14}\dyn$, and the unit of length
$1(\kms)\times\Gyr\approx1\kpc$. Viscosity and thermal diffusivity are
measured in units of $\Gyr\,\km^2\s^{-2}$. The resulting unit for
energy is $E_0$=3.2$\times 10^{50}$\,ergs, and the unit of magnetic field is
0.35$\mu$G.

We use periodic boundary conditions in all three directions for a
computational domain of size $(200\pc)^3$. Near the galactic
midplane, the cell size reported from SN-turbulence range in between
60-100\,pc \citep[e.g.][]{Korpi99}; this can be thought as the
relevant length scale for the galactic dynamo near the midplane, one
of the basic ingredients of which SN-turbulence is thought to be
\citep[see e.g.][]{Ferriere01}. Therefore, the chosen computational
domain size can be thought of as a compromise in being large enough to
contain a few turbulent SN cells but small enough to follow the phase
segregation process into cold cloudy and warm dilute phases of the
ISM.

We use the \textsc{Pencil Code},\footnote{
  \url{http://http://code.google.com/p/pencil-code/}} which is a
non-conservative, high-order, finite-difference code (sixth order in
space and third order in time) for solving the compressible
hydrodynamic equations.  Because of the non-conservative nature of the
code, diagnostics giving the total mass and total energy (accounting
for heating/cooling terms) are monitored and simulations are only
deemed useful if these quantities are in fact conserved to reasonable
precision.  The mesh spacings in the three directions are assumed to
be the same, i.e.\ $\delta x=\delta y=\delta z$. In the simulations
reported in this manuscript, the size of the computational grid is
$128^3$, yielding a resolution of roughly 1.6\,pc in our model. We
note that a resolution study in a corresponding hydrodynamic setup was
made in our previous study (BKM07), where doubling the resolution was
observed to not significantly change the flow properties.

As in our previous study (BKM07), see also \citet{Koyama04} and
\citet{Piontek05}, we include thermal conduction, which stabilizes the
gas at wavelengths larger than the critical one of the condensation
mode \citep[][]{Field65}. This wavelength is usually referred to as
the Field length, and is three orders of magnitude larger in our
simplified model than in reality (of the order of 0.001pc). The
required stabilizing effect could, in principle, be achieved by high
enough numerical diffusion arising from the scheme itself \citep[see
e.g.][]{Gazol05}, but with our high-order finite-difference scheme
this effect is not large enough with the chosen viscosity
coefficients. We emphasize that no shock or hyperviscosity has been
used in the present simulation. Therefore, the only means of
stabilizing the code is through the regular viscosity $\nu$, magnetic
diffusivity $\eta$, and thermal diffusivity $\chi$. In order to damp
unresolved ripples at the mesh scale $\delta x$ in the trail of
structures moving at speed $U$, the minimum viscosity and minimum
diffusion must be of the order of about $0.01\times U\delta x$
\citep*[see][]{BD02}; for the setup presented in this paper, the
minimal viscosity coefficients to resolve TI producing maximal
  velocities of the order of 100 km\ s$^{-1}$ with the grid and
guarantee numerical stability are $\nu$, $\chi$ and $\eta $ of $5
\times 10^{-3}$, meaning that the Prandtl numbers are of
unity. The kinetic and magnetic Reynolds numbers are defined as \EQ
\mbox{Re}= u_{\rm rms} l_f / \nu \ \ {\rm and} \ \ \mbox{Rm}= u_{\rm
  rms} l_f / \eta \EN where $l_f$ is the length scale corresponding to
the external forcing. Keeping Prandtl numbers as unity, the kinetic
and magnetic Reynolds numbers are equal.

\subsection{Diagnostic for the magnetic flux}

Instead of the commonly used point-to-point correlation of density and
magnetic field strength at each cell of the computational grid, we
propose to measure the correlation between the magnetic field strength
and density using a different procedure that cleans out the
unnecessary information of how the data points cluster about the most
probable magnetic field strength, while revealing more clearly the mean
trend in the density-magnetic field correlation. First, we divide the
logarithmic density field into $n$ bins of equal size. Then we
calculate the total magnetic flux, $\int_{S} {\bm B} \cdot dS$,
through the horizontal ($\Phi^{\rm xy}_{\rm tot}$), and the two
vertical ($\Phi^{\rm xz}_{\rm tot}$ and $\Phi^{\rm yz}_{\rm tot}$)
surfaces, and the magnetic flux through these surfaces for each
density interval ($\Phi^{\rm xy}_i$, $\Phi^{\rm xz}_i$, and $\Phi^{\rm
  yz}_i$), $i=1,...,n$. To prevent the flux diagnostic being dependent
on the amount of gas in each density bin, we need to scale with the
filling factor of each bin, $f_i$, i.e.
\begin{equation}\label{eq:diag}
\Psi^{kl}_i=\frac{\Phi^{\rm kl}_i}{f_i \Phi^{\rm kl}_{\rm tot}}.
\end{equation}
In principle, both the fluxes and the filling factors are changing as
functions of time, especially in the dynamo-active runs during the
growth phase of the magnetic field. Therefore, the flux diagnostic is
calculated only after the magnetic field growth has saturated, after
which practically no change in any of the quantities can be detected.
All the $N=\{nx,ny,nz\}$ surfaces along the three line-of-sights for
the corresponding $\{\Psi^{yz},\Psi^{xz},\Psi^{xy}\}$ are added up in
the calculation. Comparing the fluxes calculated over all the three
possible directions also allows us to detect anisotropies; this is,
however, not a relevant issue for the present study excluding
stratification, rotation and shear. This diagnostic is expected to
show a monotonically increasing trend for any system showing
correlation between the density and magnetic field, monotonically
decreasing trend for anti-correlated fields, and flat behavior, the
value approaching unity, if the quantities remain completely
uncorrelated. Before applying this diagnostic to the dynamo-generated
fields, we test it with artificially generated magnetic fields on top
of the density field from a corresponding hydrodynamic simulation (see
next section).

\begin{table*}  \begin{center} 
    \begin{tabular}{lccccccccccc} \hline
      Model &f &$\sigma$ &$u_{\rm rms}$ &$ B^{\rm tot}_{\rm rms}$ &$E_{\rm th}$ &$E_{\rm kin}$ &$E_{\rm mag}$  &Ma$_{\rm rms}$ &$\mbox{Rm}$ &$\lambda^{-1}_{\rm B}$ &$\Delta \rho$\\
      & & &[kms$^{-1}$] &[$\mu$G] &[$E_0$] &[$E_0$] &[$E_0$] & &$(\mbox{Re})$ &[Myrs] &\\
      \hline \hline
TSf   &7  &1 &0.891 &-     &0.943 &0.005 &-         &0.100 &12  &-   &2\\
TSe   &10 &1 &1.210 &0.231 &0.944 &0.010 &0.002     &0.135 &16  &500 &2 \\
HDTSa &20 &1 &2.453 &-     &0.945 &0.038 &-         &0.271 &33  &-   &3\\
TSa   &20 &1 &2.054 &0.925 &0.946 &0.028 &0.028     &0.230 &27  &89  &4\\
TSanh &20 &1 &2.357 &-     &0.946 &0.036 &-         &0.264 &31  &-   &3\\
TSd   &35 &1 &3.384 &1.500 &0.948 &0.074 &0.064     &0.377 &49  &45  &10\\
HDTSb &50 &1 &5.648 &-     &0.945 &0.186 &-         &0.617 &75  &-   &20\\
TSb   &50 &1 &4.786 &2.051 &0.952 &0.147 &0.137     &0.527 &64  &37  &27\\
TSbnh &50 &0 &5.142 &-     &0.951 &0.171 &-         &0.571 &69  &-   &10\\
HDTSc &70 &1 &7.762 &-     &0.943 &0.342 &-         &0.848 &103 &-   &97\\
TSc   &70 &1 &6.677 &2.455 &0.958 &0.263 &0.203     &0.703 &89  &27  &52\\
TScnh &70 &0 &6.812  &-     &0.954 &0.294 &-        &0.750 &91  &-    &50\\
\hline
SSe    &10 &1 &2.118 &-     &0.387 &0.018 &-        &0.300 &28 &-  &371\\
SSf    &15 &1 &3.400 &0.189 &0.428 &0.034 &0.002    &0.392 &45 &510 &390\\
HDSSa  &20 &1 &3.299 &-     &0.433 &0.042 &-        &0.434 &44 &-  &431\\
SSa    &20 &1 &3.067 &0.448 &0.447 &0.021 &0.007    &0.289 &52 &200&490\\
SSanh  &20 &0 &2.910 &-     &0.436 &0.029 &-        &0.351 &39 &-  &461\\
SSd    &35 &1 &4.345 &0.787 &0.494 &0.023 &0.022    &0.282 &58 &69 &482\\
HDSSb  &50 &1 &6.023 &-     &0.564 &0.082 &-        &0.528 &80 &-  &973\\
SSb    &50 &1 &4.621 &1.190 &0.558 &0.043 &0.048    &0.371 &62  &55 &792\\
SSbnh  &50 &0 &5.466 &-     &0.552 &0.067 &-        &0.467 &73  &-  &633\\
HDSSc  &70 &1 &8.114 &-     &0.630 &0.145 &-        &0.679 &108 &   &1365\\
SSc    &70 &1 &5.488 &1.666 &0.628 &0.067 &0.094    &0.422 &73  &33 &482\\
SScnh  &70 &0 &7.995 &-     &0.634 &0.117 &-        &0.583 &97  &-  &1315\\
\end{tabular}
\caption{Results from various runs. The thermally stable runs are
  labelled with 'TS', thermally unstable with 'SS', purely hydrodynamic
  runs with 'HD', and non-helical with 'nh'.}\label{table:runs}
\end{center}
\end{table*}

\section{Results}

The produced runs, their basic parameters, and some ensemble-averaged
quantities describing the basic results are listed in
Table~\ref{table:runs}. Rms velocity $u_{\rm rms}$, rms total magnetic
field $B^{\rm tot}_{\rm rms}$, thermal $E_{\rm th}$, kinetic
$E_{\rm kin}$, magnetic $E_{\rm mag}$ energies, rms Mach number
$Ma_{\rm rms} = u_{\rm rms}/c_{\rm rms}$, where $c_{\rm rms}$ is the
rms sound speed, and magnetic Reynolds number $\mbox{Rm}$, equal to
$\mbox{Re}$, are all calculated as averages over the whole volume and
over several Gyrs of time after the saturated state has been
reached. For the hydrodynamic runs, instead of the magnetic Reynolds
number, the fluid Reynolds number is given.  The growth rate,
$\lambda_{\rm B}$, is measured from $B^{\rm tot}_{\rm rms}$
during the exponential phase of the magnetic field growth; in the
table, we list its inverse to illustrate the growth time scale of
the magnetic field in Myrs. To measure the density contrast, $\Delta
\rho$, averages of the extremum values of densities over the saturated
state are used. The labelling system of the runs is as follows: 'HD'
refers to hydrodynamic runs excluding magnetic field; 'TS' refer to
the usage of thermally stable cooling function; 'SS' refers to the
thermally unstable cooling function; 'nh' refers to the usage of a
non-helical forcing function.

\begin{figure}[!t]
  \centering \includegraphics[width=0.98\columnwidth]{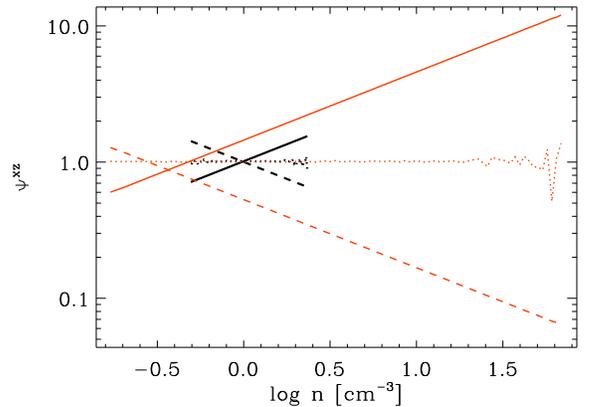}
  \caption{Diagnostic Eq.~(\ref{eq:diag}) calculated by imposing an
    artificially generated magnetic field on top of the thermally
    stable hydrodynamic Run~HDTSa (black lines) and thermally unstable
    Run~HDSSa (orange/grey lines). Different types of magnetic field
    dependencies on density are plotted with different line styles:
    solid line, $|\bm{B}|=B_0 \rho^{1/2}$, dashed line,
    $|\bm{B}|=B_0 \rho^{-1/2}$, dotted line: random magnetic field.}
  \label{fig:art_corr}
\end{figure}

First we demonstrate the operation of the diagnostic defined in
Eq.~(\ref{eq:diag}), using some purely hydrodynamic calculations
otherwise identical to the magnetohydrodynamic ones, but excluding the
magnetic field. We use the density fields from the thermally
stable run HDTSa and its thermally unstable counterpart HDSSa, on top
of which we artificially add a magnetic field directed in the positive
$y$-direction. The amplitude of the field is given either by (a)
$|{\bm B}|=B_0\rho^{1/2}$, (b) $|{\bm B}|=B_0\rho^{-1/2}$ or (c) the
amplitude of the field is a random number in the range $[0,B_0]$, with
$B_0$ fixed to unity. In such a setup, only the flux component
$\Psi^{xz}$ is nonzero, and this quantity is plotted for the different
cases in Fig.~\ref{fig:art_corr}. As evident from this figure, for a
setting with a positive correlation between the density and magnetic
field strength (case a; plotted with solid lines of varying color),
the flux diagnostic shows a monotonically increasing trend. For an
anti-correlated setup (case b; plotted with dashed lines of varying
color), the diagnostic shows a monotonically decreasing trend. For the
case of completely uncorrelated magnetic field strength and density
(case c; plotted with dotted lines of varying color) the curve is flat
and has a value around unity. The density range in the
thermally unstable Run~HDSSa is larger than in the stable counterpart,
Run~HDTSa, but the diagnostic behaves similarly in both cases.

\begin{figure*}[t!]\begin{center}
\includegraphics[width=0.24\textwidth]{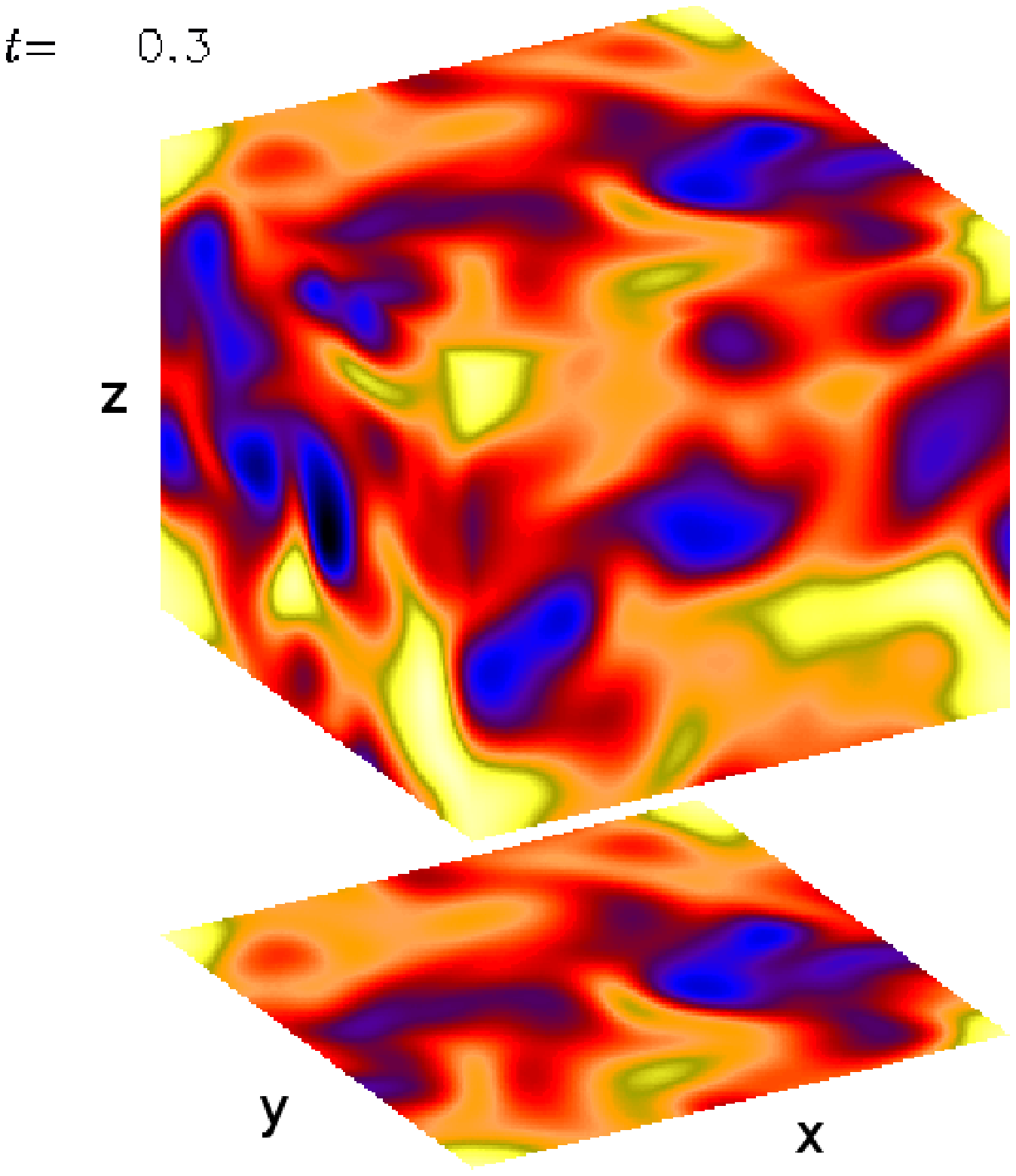}\includegraphics[width=0.24\textwidth]{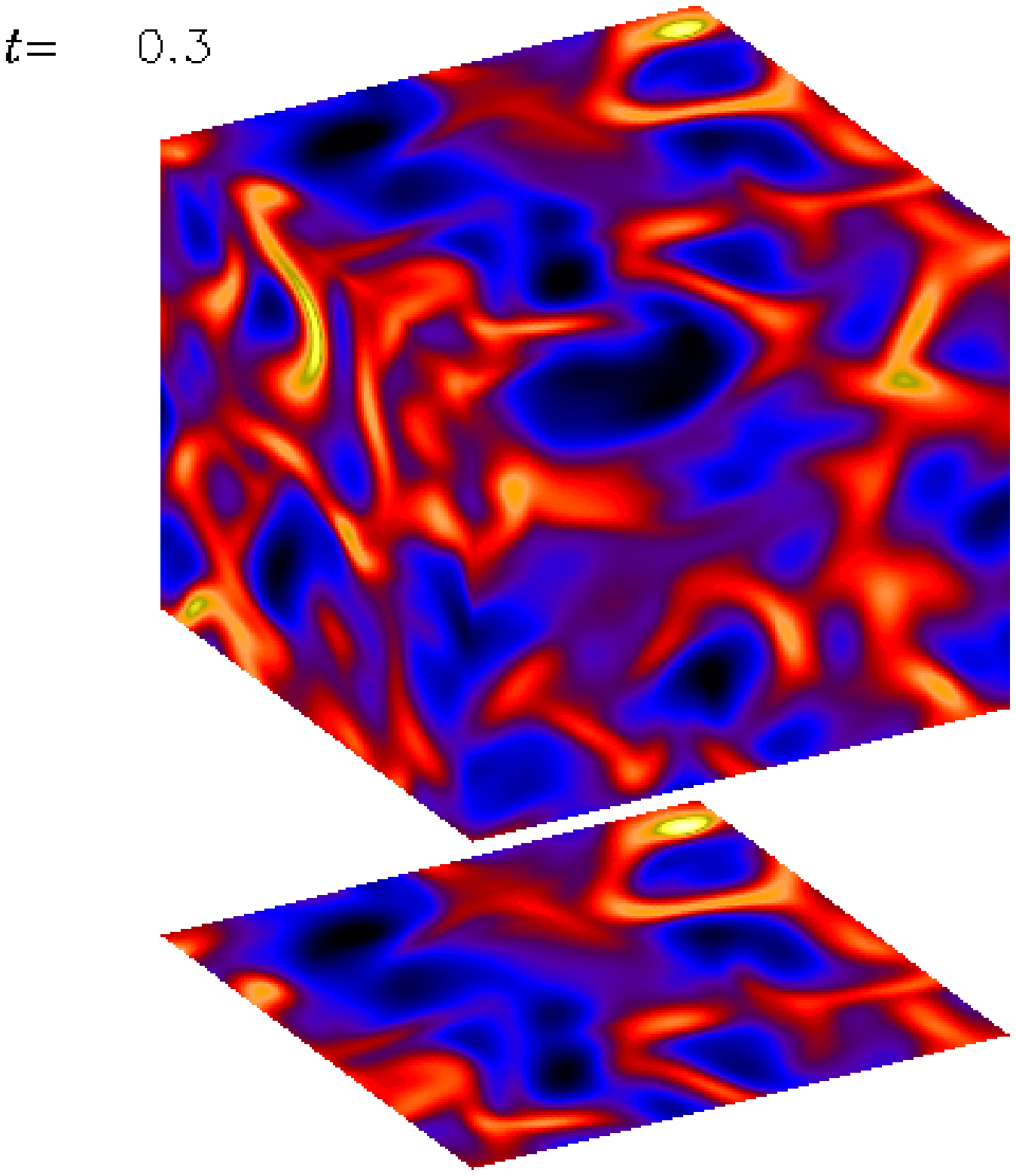}\includegraphics[width=0.24\textwidth]{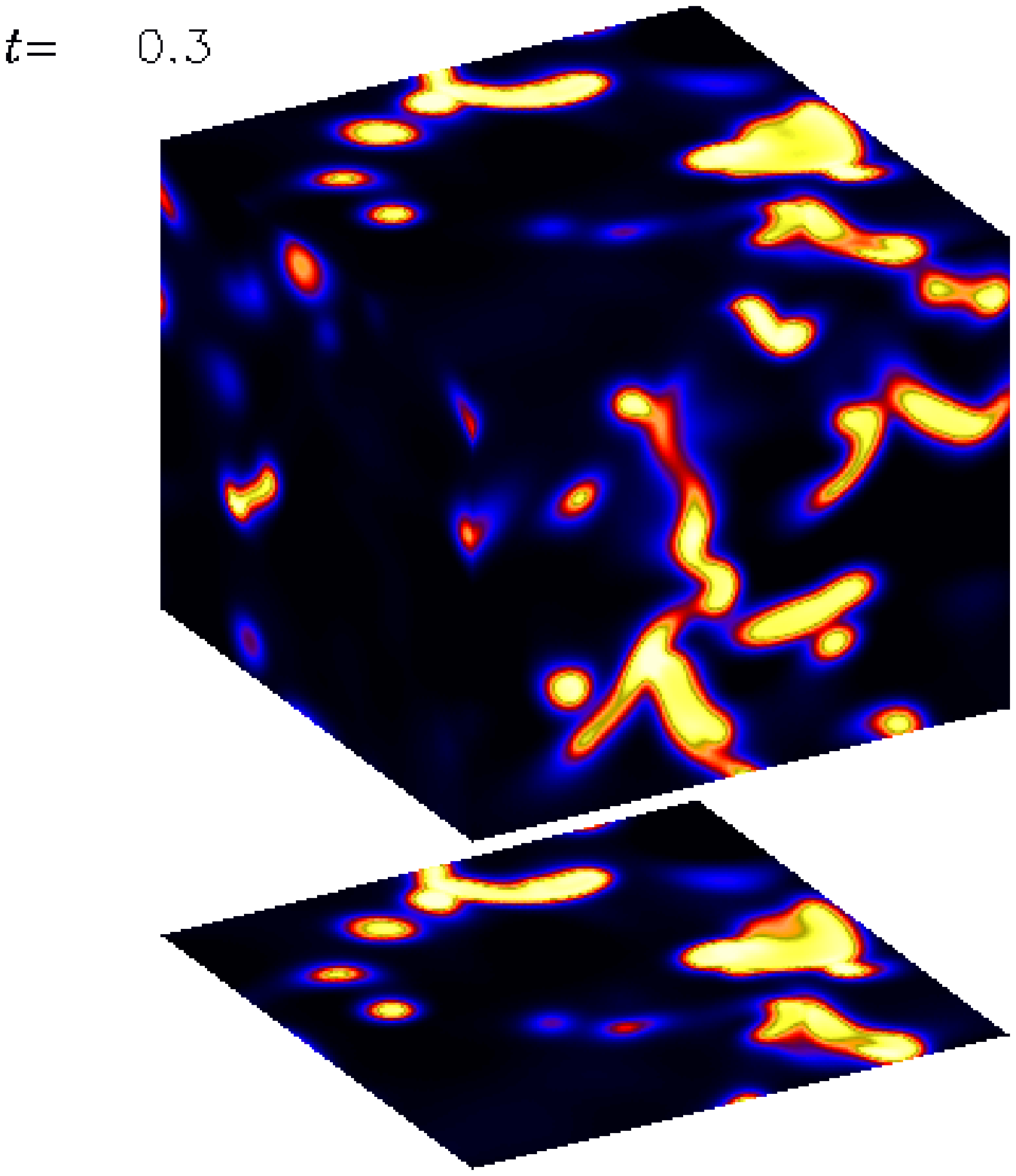}\includegraphics[width=0.24\textwidth]{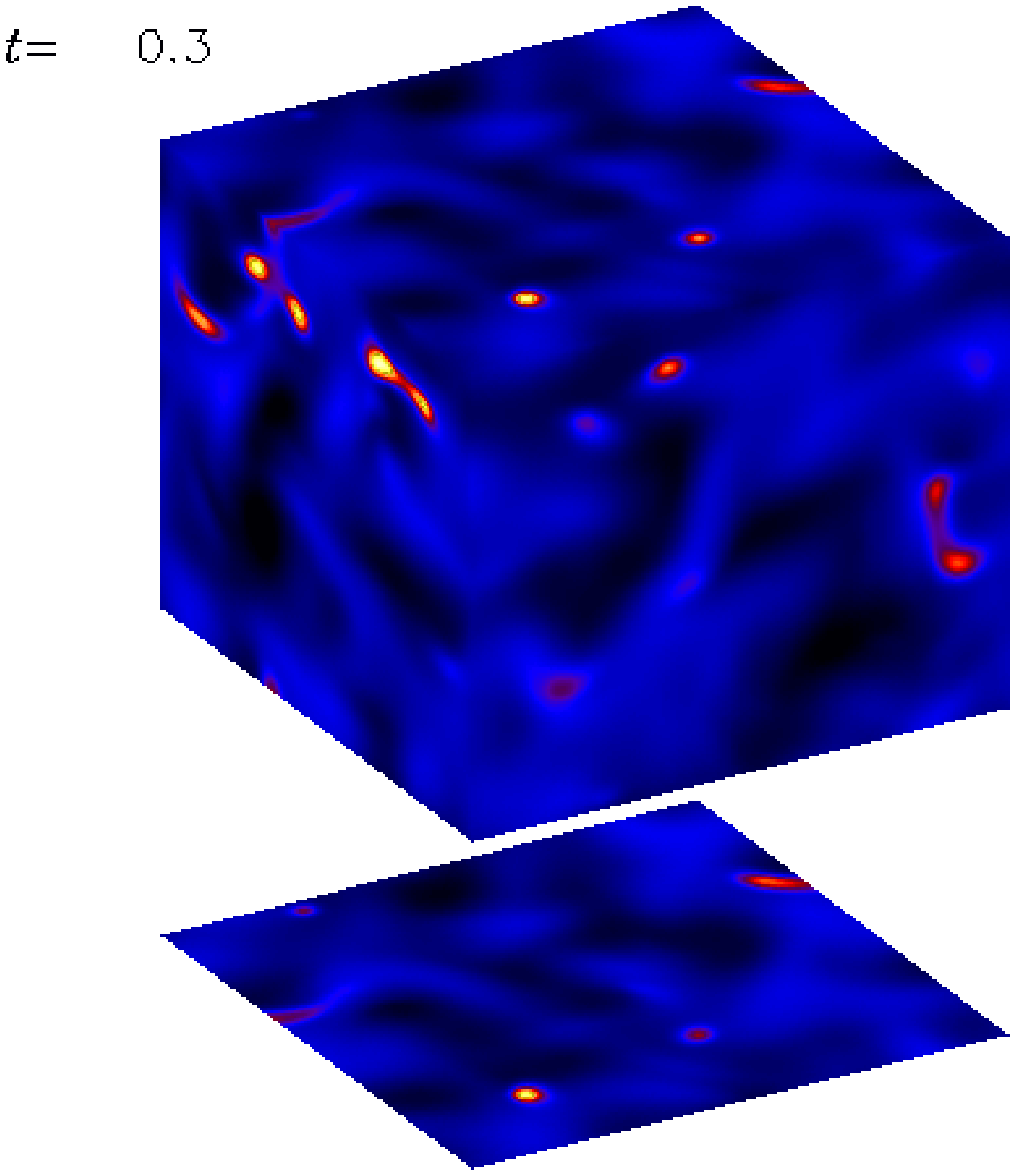}
\includegraphics[angle=90,width=0.02\textwidth]{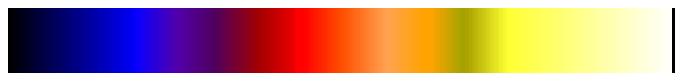}
\end{center}\caption[]{Snapshots of the logarithmic density fields
  from different hydrodynamic runs at 250 Myrs. From left to right:
  HDTSa, HDTSc, HDSSa, HDSSc.}\label{hydro_rsnap}\end{figure*}

\begin{figure*}[t!]\begin{center}
\includegraphics[width=0.33\textwidth]{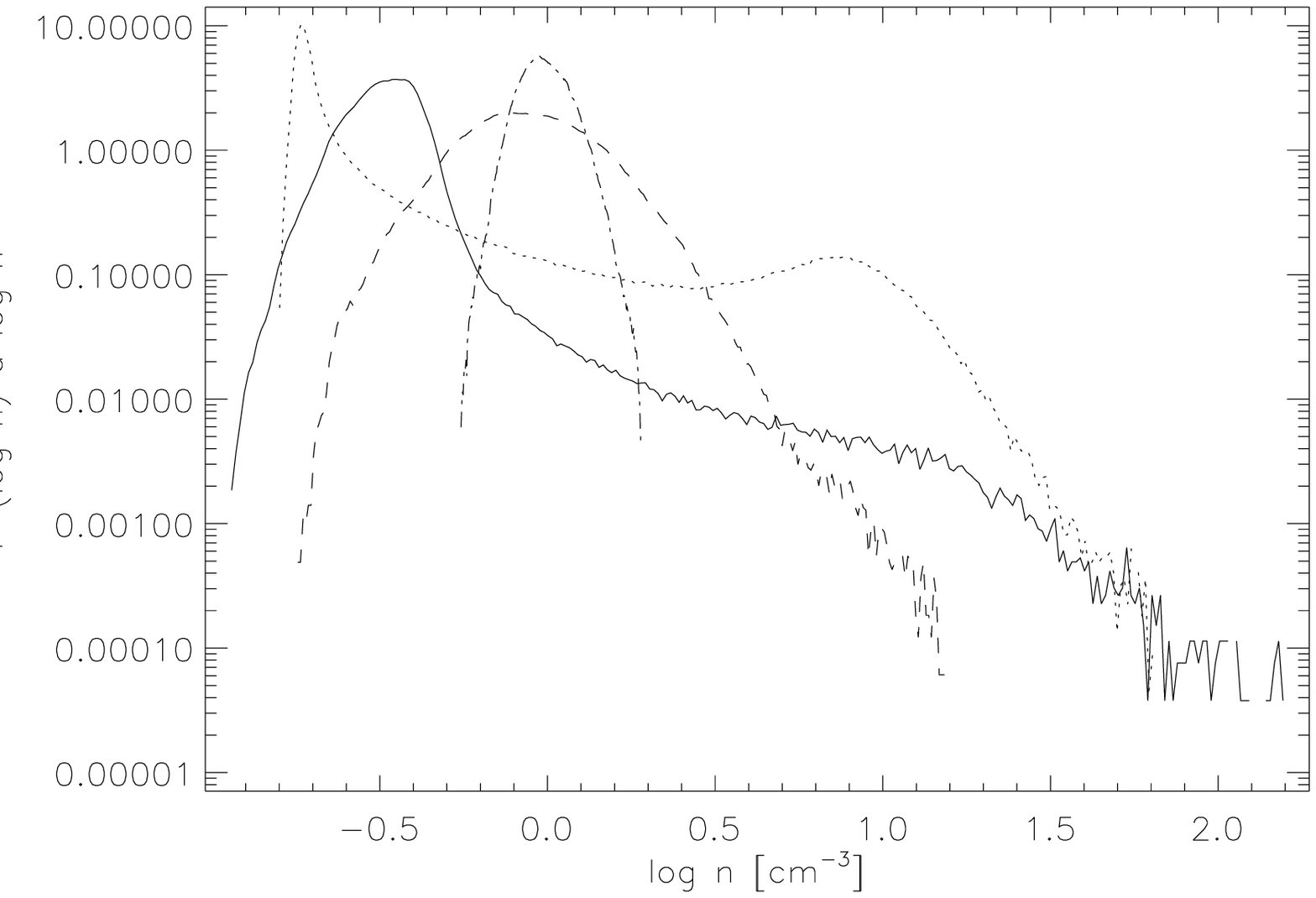}\includegraphics[width=0.33\textwidth]{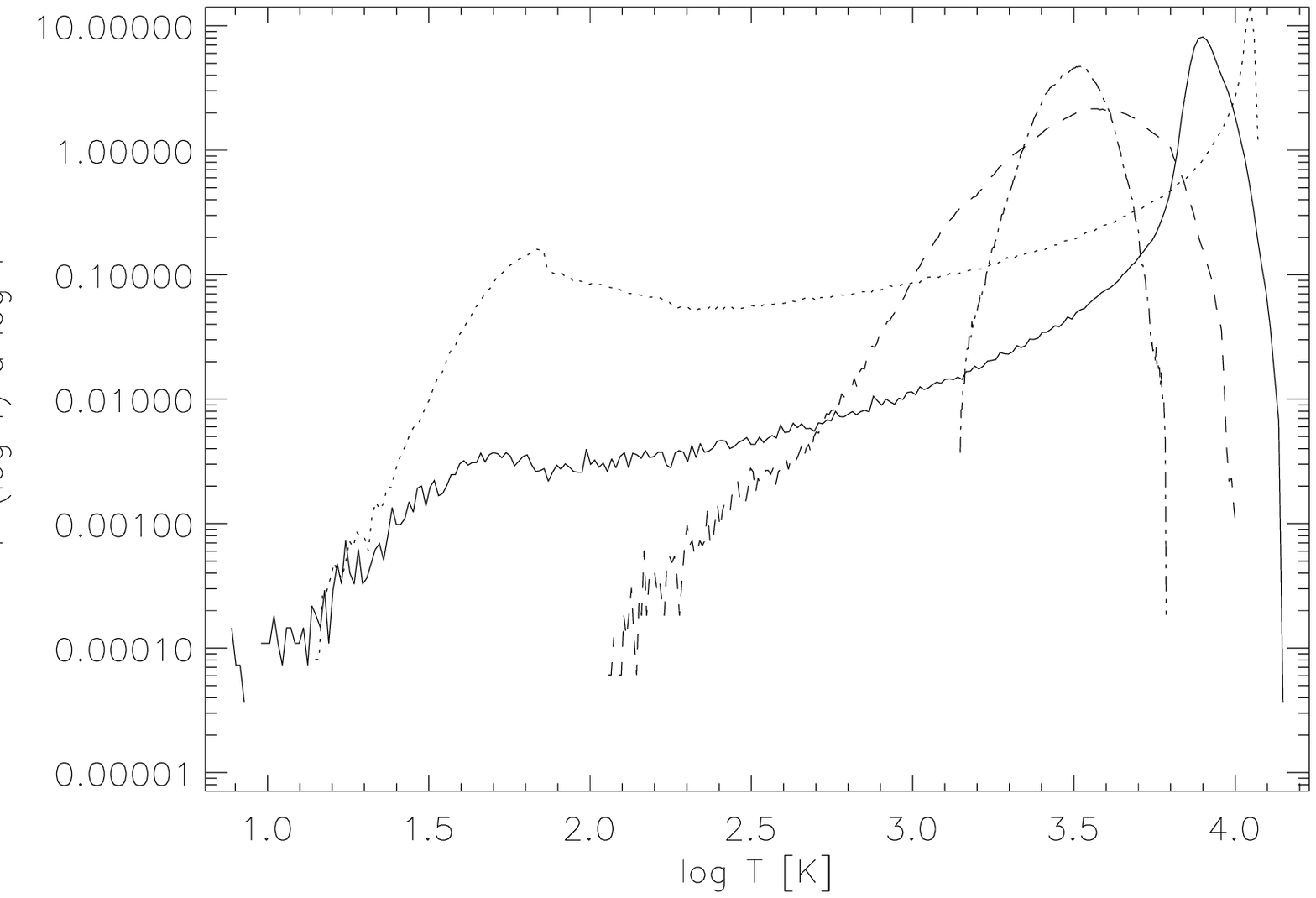}\includegraphics[width=0.33\textwidth]{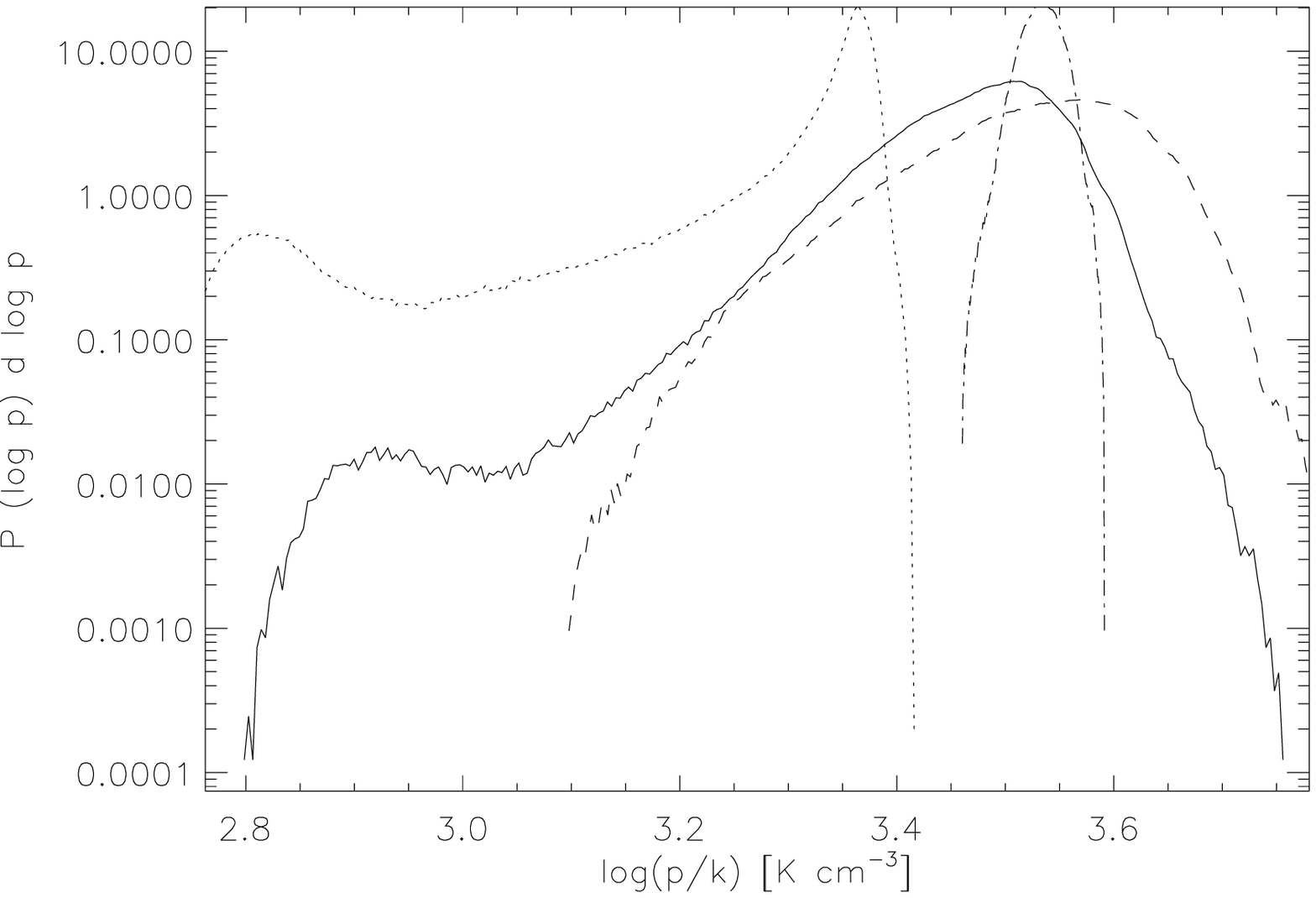}
\end{center}\caption[]{Probability density functions for the thermally
  unstable hydrodynamic runs HDSSc (solid line) and HDSSa (dotted
  line), and the thermally stable hydrodynamic runs HDTSc (dashed
  line) and HDTSa (dashed-dotted line) at 250 Myrs of the simulation
  runs. On the left we show the PDF of logarithmic density, in the
  middle the PDF of logarithmic temperature, and on the right the PDF
  of logarithmic pressure.}\label{hydro_pdfs}\end{figure*}

\begin{figure*}[t!]\begin{center}
\includegraphics[width=0.24\textwidth]{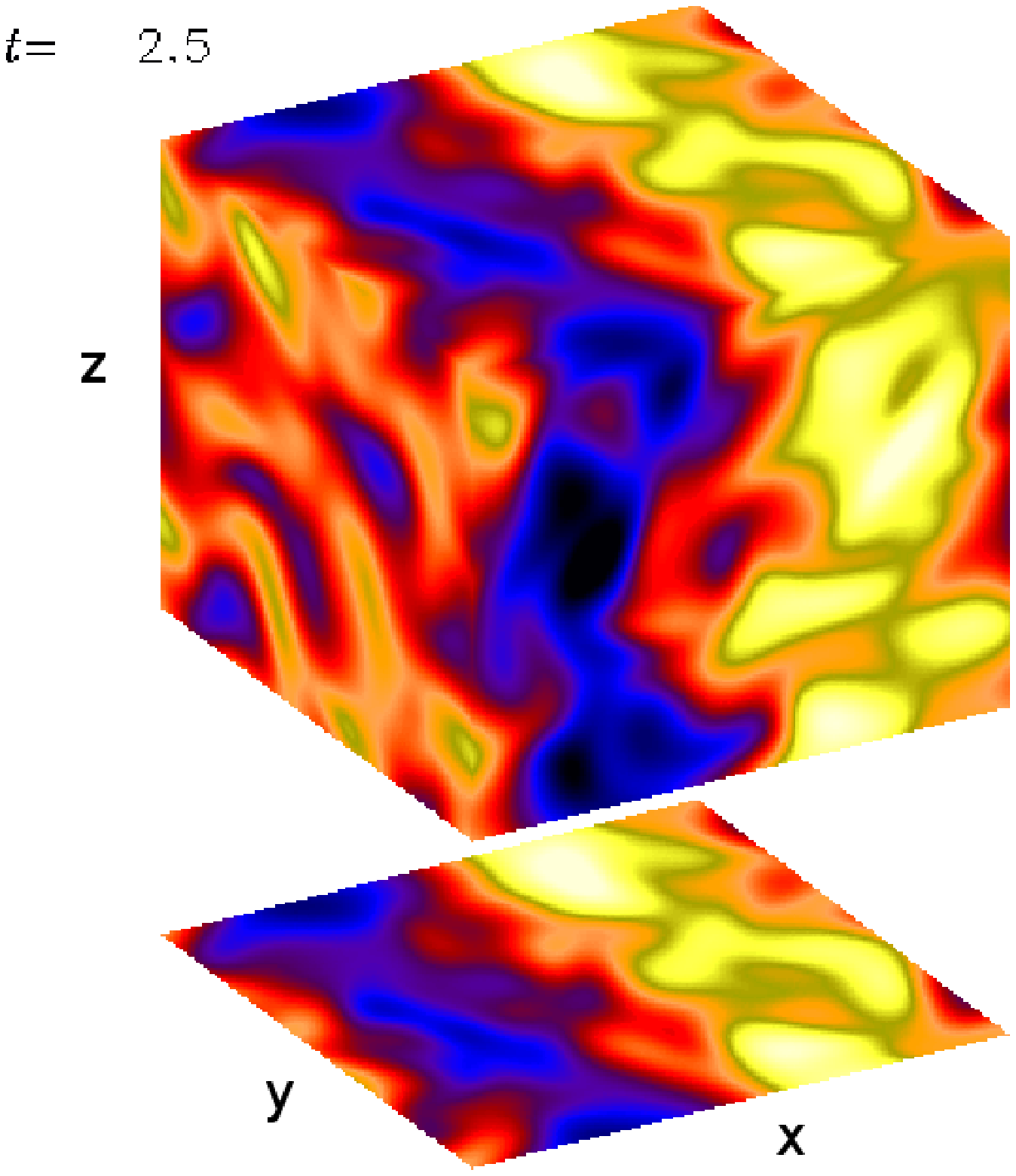}\includegraphics[width=0.24\textwidth]{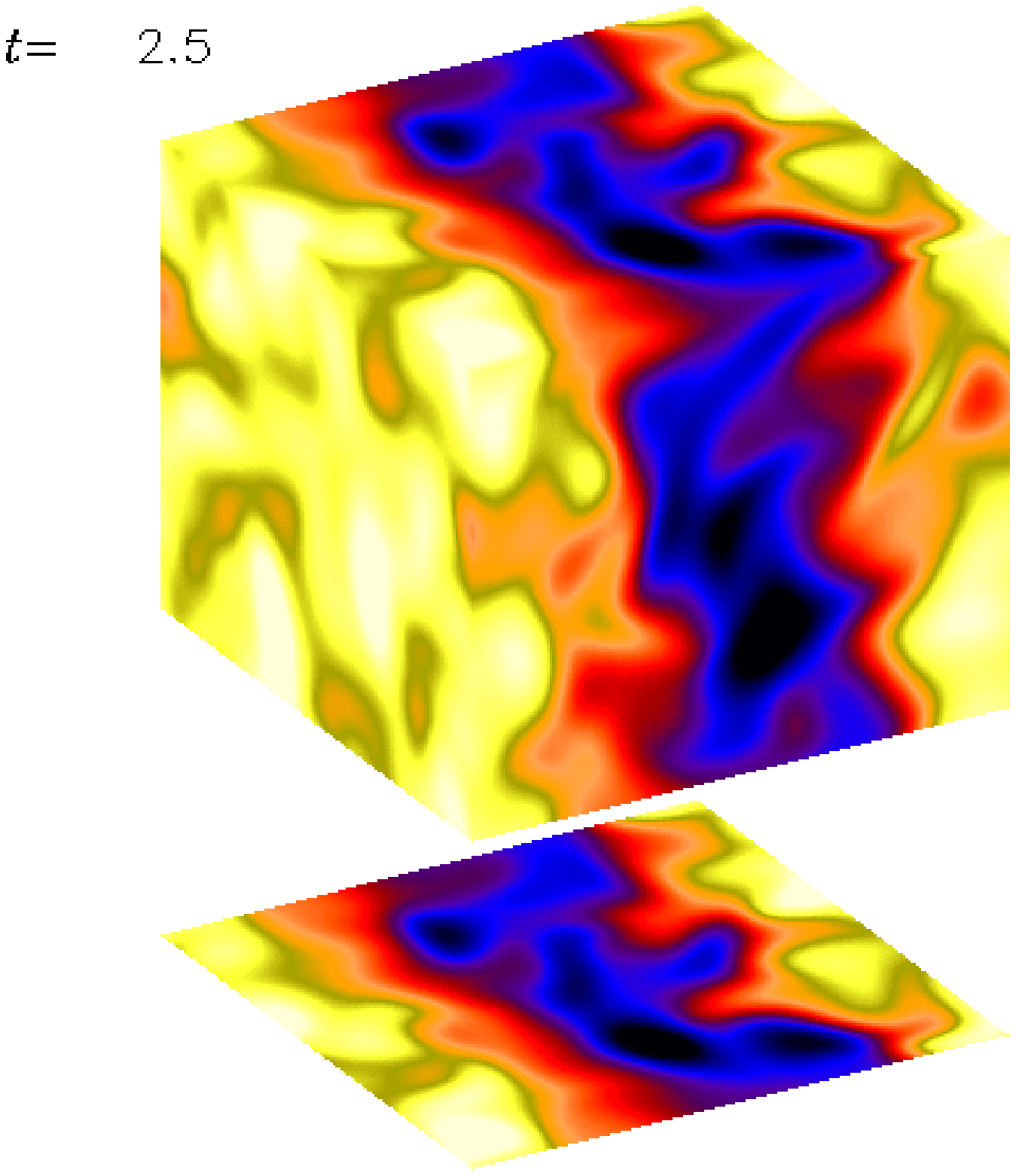}\includegraphics[width=0.24\textwidth]{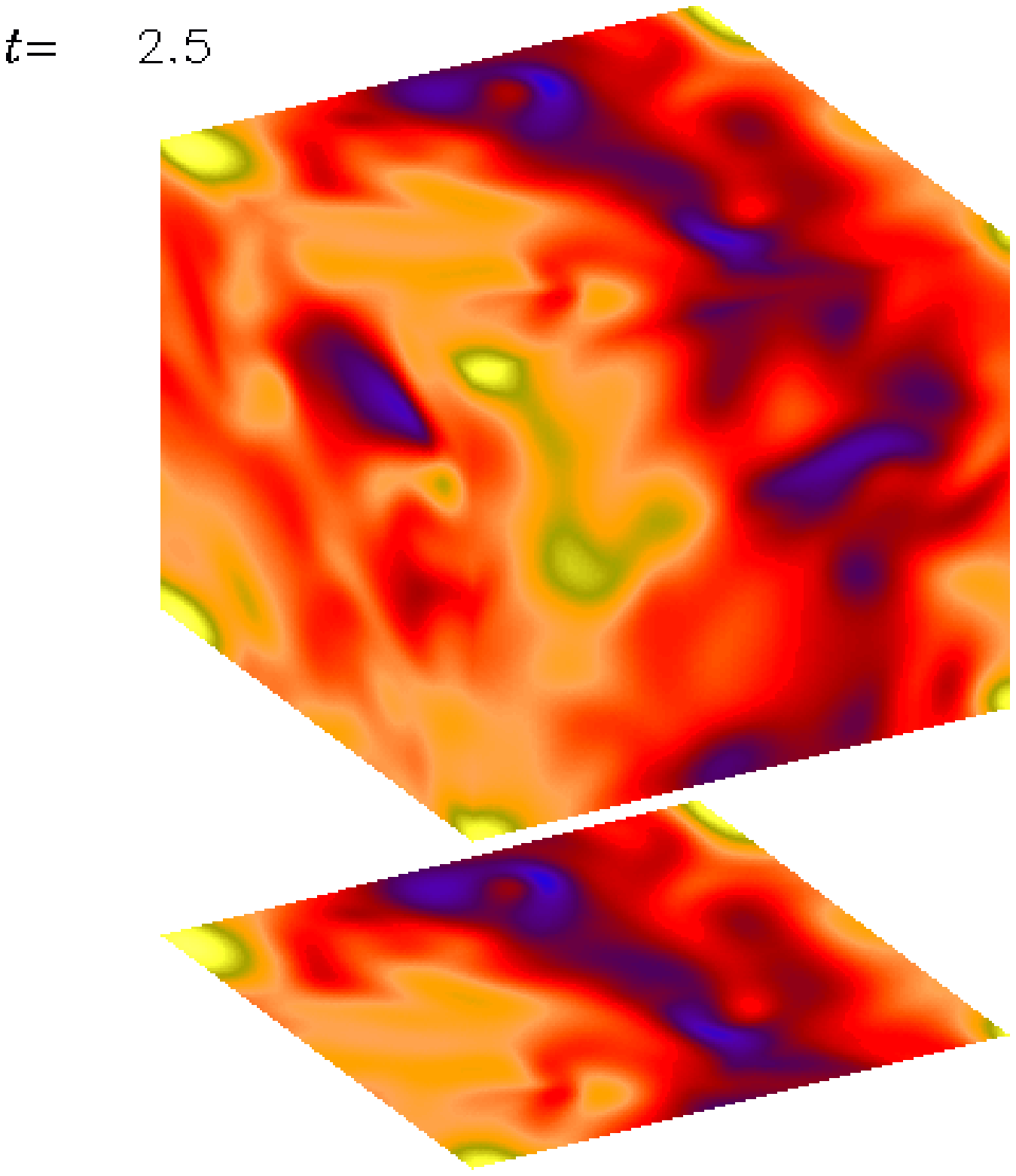}\includegraphics[width=0.24\textwidth]{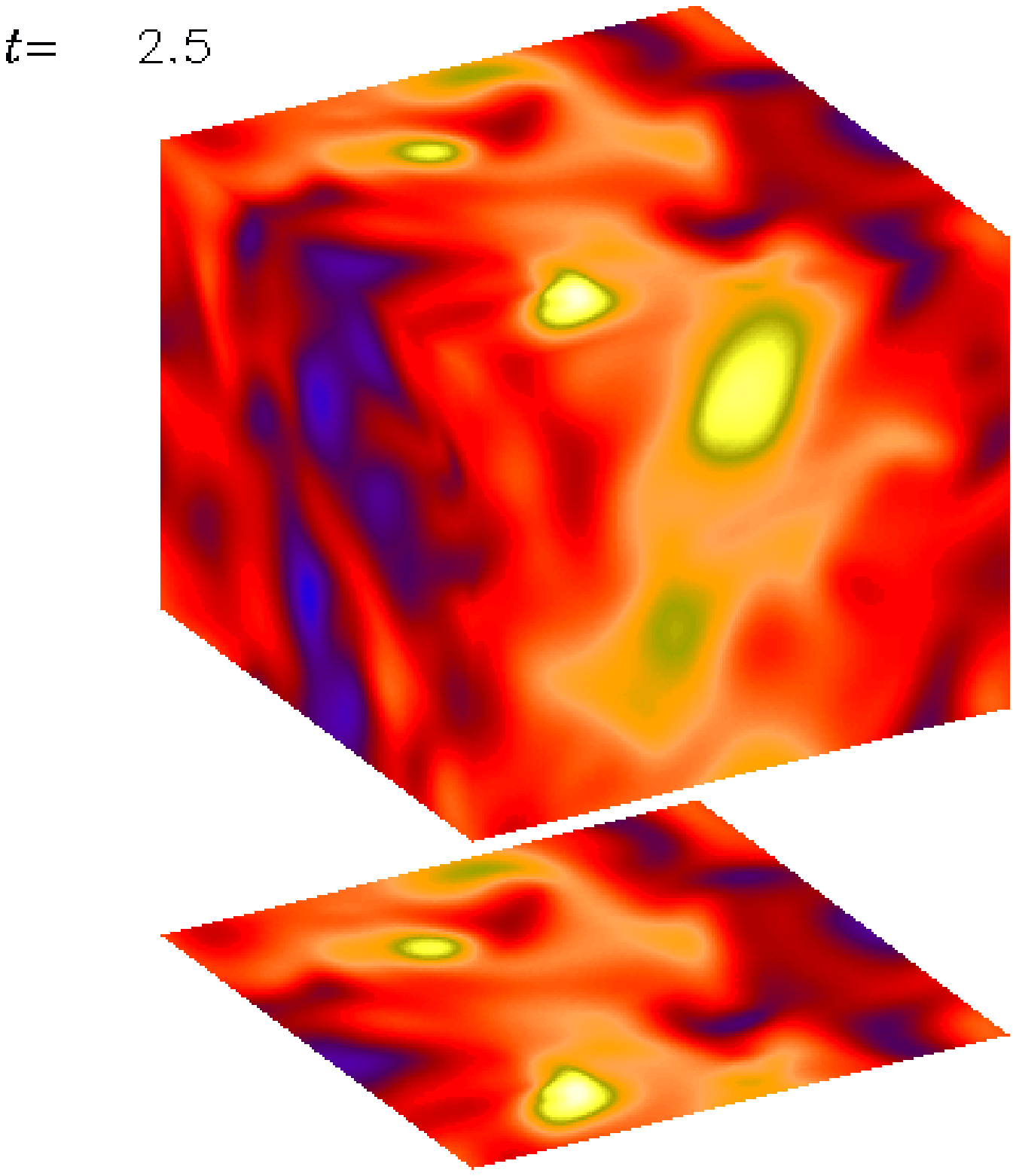}
\includegraphics[angle=90,width=0.02\textwidth]{fig10.eps}
\end{center}\caption[]{Snapshots of the magnetic field strength from
  the thermally stable run TSa and thermally unstable run SSa at 2.5
  Gyrs of evolution. From left to right: $B_y$ in TSa, $B_z$ in TSa, $B_y$ in
  SSa, and $B_z$ in SSa. The minimum value plotted with blue color
  corresponds to -1.2\,$\mu$G and the maximum value to 1.2\,$\mu$G,
  used in all plots.}\label{bsnap}\end{figure*}

\subsection{Hydrodynamic state of the system}

Let us first compare the purely hydrodynamic runs with different
cooling functions and forcings. For the runs with thermally stable
cooling functions (labeled with HDTS in Table~\ref{table:runs}), the
formation of density structures occurs solely by the mixing and
compression caused by the turbulent flow. For cases with the weakest forcing,
Run~HDTSa, rms-velocity of the order of 2\,kms$^{-1}$, and density
contrast of 3 is observed. The density field of this run at 250\,Myrs
is illustrated in Fig.~\ref{hydro_rsnap} leftmost panel, from which
it can easily be seen that the turbulence is so weak that the
density structures generated by the flow do not show large contrasts
and remain quite large in size. When the strength of the forcing is
increased, the density contrast gets larger, as expected, so that in
Run~HDTSb, with 2.5 times stronger forcing, the contrast is roughly
20, and in Run~HDTSc, with 3.5 times higher forcing, roughly 100,
the respective rms-velocities being 6 and 8\,kms$^{-1}$. The
appearance of the flow field, illustrated in Fig.~\ref{hydro_rsnap},
the second panel from the left for Run~HDTSc, has somewhat changed,
the density structures generated now showing up as narrow
filaments. 

The effect of increasing forcing is also apparent in the probability
distribution functions (hereafter PDFs) of the quantities, plotted in
Fig.~\ref{hydro_pdfs} for Run~HDTSa (dashed-dotted lines) and
Run~HDTSc (dashed lines). Evidently, the distributions of all the
quantities are all unimodal, with the width of them increasing as the
forcing is increased. The density PDF is clearly reflecting the
somewhat increased capability of the flow to generate denser
structures with a smaller filling factor. The denser the structures the
flow is capable of creating, the more will they cool, as the cooling
term is of the form $\rho \Lambda$, modifying the rather symmetrical
temperature PDF for Run~HDTSa to exhibit an extended wing towards
lower temperatures for Run~HDTSc. This excess cooling will aid the
dense regions in becoming denser, and a similar asymmetrical wing
towards higher densities is seen in the density PDF for
Run~HDTSc. This is in agreement with the results of \citet*{PV98}, who
studied highly compressible polytropic flows with a varying Mach
number. For polytropic indeces less than one, a very similar wing,
approaching power-law behavior, was seen to develop towards higher
densities. Similar, skewed density PDFs towards higher densities have
also been reported from 3D-simulations of polytropic flows by
\citet*{NP99}, interpreted to be caused by the effective $\gamma$
being reduced to below one in the regions of high compression.

For the runs with thermally unstable cooling functions (labeled with
HDSS in Table~\ref{table:runs}), a major part of the formation of
density structures is expected to occur due to the thermal instability,
although the turbulence can also be expected to alter this picture
considerably \citep[see e.g. BKM07;][]{Gazol05}. As can be seen
from the two rightmost panels of Fig.~\ref{hydro_rsnap}, the
appearance of the flow is very different from the thermally stable
cases (the two leftmost panels in that figure). The action of the
thermal instability leads to a very rapid (timescale of few tens of
Megayears) segregation of the gas into cold cloudy and warm diffuse
phases, the dense clouds clearly visible in the snapshots of the
density field. Similarly to the thermally stable cases, the increasing
vigor of turbulence can be seen to enhance the condensation process, so
that excess cooling due to turbulent compression leads to an even larger
density contrast and smaller structures. 

We also plot the PDFs of the quantities for Run~HDSSc (solid lines)
and Run~HDSSa (dotted lines) in Fig.~\ref{hydro_pdfs}. For the
Run~HDSSa with relatively weak forcing, the density and temperature
distributions are still clearly bimodal, as has commonly been observed
for thermally unstable flows, e.g. by BKM07, although large amounts of
gas can be seen in the 'forbidden' regime of thermally unstable
temperatures with $\beta < 1$. For the Run~HDSSc, the bimodality of
the density and temperature PDFs is much less pronounced; both
distributions have become wider, and the peaks have been shifted
towards higher densities and lower temperatures. This is reflected
also in the peak of the pressure distribution. For the strongest
forcing, the most probable pressure is somewhat higher than for the
weaker forcing. A similar trend in the pressure distribution can also
be seen in the thermally stable runs. The filling factor of the cold
component gets considerably smaller as the forcing is increased,
especially clearly seen when comparing the two leftmost panels of
Fig.~\ref{hydro_rsnap}. This is also clearly manifested in the PDFs
showing decreasing significance of the distribution of the cold
phase. Due to the skewness of the density distribution towards higher
densities, also the pressure distribution was asymmetric in the
thermally stable case. In the thermally unstable cases, the pressure
PDF is also clearly asymmetric, but in addition exhibits an extended
wing towards the low pressures, and this wing gets more pronounced
with increasing forcing amplitude, visible in the rightmost panel of
Fig.~\ref{hydro_pdfs}. In our study, the vigor of forcing is still
modest and the resulting Mach numbers are reasonably low in comparison
to, e.g. \citet{Gazol05}, who studied a similar system in both the
sub- and supersonic regimes. In their results, the signatures of TI in
the PDFs are much less pronounced; on the other hand, the extended
wing seen to develop in our pressure PDFs, is not reported in theirs.

As evident from Table~\ref{table:runs}, even though the cooling
functions were scaled to produce the same integrated cooling over the
whole temperature range for constant density, the thermal energy
content in the thermally unstable runs is roughly half of that of the
thermally stable counterparts. Obviously, this difference is due to
the fact that the density is not constant, but actually very different
in the thermally stable versus thermally unstable systems, as was
discussed in length in this section, causing a nonlinear back-reaction
to the cooling process, significantly enhancing its efficiency.
For the thermally stable runs, however, the thermal energy is almost
constant with any value of forcing investigated. For thermally
unstable cases, the thermal energy is an increasing function of
forcing.

\subsection{Non-helical forcing}

\begin{figure}[!ht]\begin{center}
    \includegraphics[width=0.98\columnwidth]{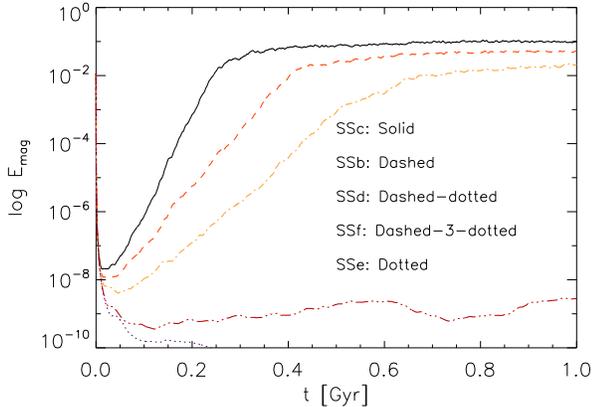}
    \caption[]{Time evolution of the magnetic fields in some of the
      thermally unstable runs.}\label{fig:emag}\end{center}\end{figure}

Using the same setup as for the hydrodynamic runs, we additionally
include a very small random seed field of rms amplitude $B_{\rm
    init}=4 \times 10^{-4}\mu$G and zero mean to the system, and
follow the evolution of the system with the full set of MHD-equations
with both cooling functions. In the following, we divide the
magnetic field into its large-scale and fluctuating components,
i.e. ${\bm B}^{\rm tot}={\bm B} + \bm{b}$, the condition
$\left<\bm{b}\right>$=0 holding. In the runs TSanh-TScnh (thermally
stable) and SSanh - SScnh (thermally unstable) the forcing function
used has no relative helicity, i.e. $\sigma$=0. Large-scale dynamo
action, i.e. the growth of the large-scale magnetic field $\bm{B}$
that has a non-zero mean, has been found in systems with non-helical
forcing when large-scale shear has been included
\citep[][]{Yousef08}. Systems such as the one studied here, however,
can be expected to develop only small-scale dynamo action, i.e. the
growth of the fluctuating component $\bm{b}$, when subjected to
non-helical forcing. The results for the six non-helical runs produced
are listed in Table~\ref{table:runs}. The forcing amplitude was
increased as high as was permissible by the numerical scheme without
invoking shock-capturing viscosities. No small-scale dynamo action was
observed, i.e. the total magnetic field strength was exponentially
decaying identically to the helical cases without large-scale dynamo
action (see Fig.~\ref{fig:emag}), in any of the runs, i.e. in the
range of Reynolds numbers $31-97$. This indicates that in the system
under investigation, the critical Reynolds number for small-scale
dynamo action is larger than the ones reached in our runs, and that if
magnetic fields are generated in the helical cases investigated, they
arise solely due to the action of a large-scale dynamo.

\cite{HBM04} studied the onset of dynamo action in an isothermal
system with non-helical forcing. As in this study, the vigor of
forcing was increased, and it was observed that the critical Reynolds
number for the onset of the small-scale dynamo was increasing with the
Mach number. Shock-capturing viscosities were used to reach the trans-
and supersonic regimes. For subsonic flows, the critical Reynolds
number for small-scale dynamo action was of the order of 35, while for
supersonic flows with Mach numbers exceeding unity, the critical
Reynolds number was almost doubled up to 70. Our weakest forcing cases
with modest Mach numbers have Reynolds numbers slightly below (TSanh,
31) or slightly exceeding the limit (SSanh, 39) found by \cite{HBM04},
and in the light of their results, it is understandable that no
small-scale dynamo action can be seen. The modest and high amplitude
forcings used here do not yet produce Mach numbers even close to the
supersonic limit, but their Reynolds numbers are clearly larger than
the critical limits found by \cite{HBM04}. Still, no small-scale
dynamo action is seen, indicating that the inclusion of thermodynamics
with the quite complex heating and cooling properties further alters
the Mach-number dependence of the critical Reynolds number for
small-scale dynamo action.

As can be seen from Table~\ref{table:runs}, the non-helical runs
produce somewhat weaker turbulence than the runs with purely helical
forcing $\sigma$=1. This is also reflected by the density
  distribution, so that the density contrast is reduced in the
  non-helical runs compared to the helical counterparts. This is
manifested by smaller rms values for velocity, and thereby also
smaller Reynolds numbers. The differences in the Reynolds numbers are
not very large, so we believe that our conclusion about the absence of
small-scale dynamo in the helical runs still holds.

\begin{table*}  \begin{center} 
    \begin{tabular}{lcccccccc} \hline
      Model &f &h &$\sqrt{\left<B_x^2\right>_{xz}}$ &$\sqrt{\left<B_x^2\right>_{xy}}$ &$\sqrt{\left<B_y^2\right>_{yz}}$  &$\sqrt{\left<B^2_y\right>_{xy}}$ &$\sqrt{\left<B^2_z\right>_{yz}}$ &$\sqrt{\left<B^2_z\right>_{xz}}$ \\
      \hline \hline
TSe    &10 &1 & - & 0.06 &0.05 & 0.07 &0.05 & - \\
TSa    &20 &1 & - & - &0.49 & - &0.43 & - \\
TSd    &35 &1 &0.70 & - & - & - & - & 0.79 \\
TSb    &50 &1 &0.98 & - & - & - & - & 1.01 \\
TSc    &70 &1 & -  & 1.18 & - & 1.23 & - & - \\
\hline
SSf    &15 &1 & - & 0.09 &- & 0.09 &- & - \\
SSa    &20 &1 & - & - &0.21 & - &0.23 & - \\
SSb    &50 &1 &0.58 & - & - & - & - & 0.60 \\
SSc    &70 &1 & - & 0.80 & - & 0.81 & - & - \\
\end{tabular}
\caption{The rms-stregth of the $k=1$-mode, measured in $\mu$G, generated in the
  runs.}\label{table:mgfs}
\end{center}
\end{table*}

\subsection{Thermally stable magnetohydrodynamic runs}

Next we repeat the magnetohydrodynamic runs with helical forcing,
starting with the thermally stable cooling function, which is expected to
produce large-scale dynamo action \citep[e.g.][]{Axel01}. The Run~TSf
with the forcing amplitude $f=7$ is sub-critical to dynamo action,
i.e. the small seed field exponentially decays. The Run~TSe with
$f=10$, however, produces a very slowly growing magnetic field, with
the time scale of growth of roughly $500$\,Myrs. The growth of the
field saturates only after 4 Gyrs, the saturation energy being less
than half of the kinetic energy of turbulence. Inspecting the Reynolds
numbers realized in Runs~TSe and TSf from Table~\ref{table:runs}, the
critical Reynolds number for the large-scale dynamo in this system,
therefore, is roughly between 12-16.

Runs with higher forcing (Runs~TSa, TSd, TSb and TSc), with a forcing
ranging between 20-70, all produce dynamo action that leads to
the generation of magnetic fields, the energy of which is in rough
equipartition with the kinetic energy of turbulence for intermediate
forcings (TSa and TSd), but slightly lower than the equipartition
value for the highest forcings investigated (TSb and TSc). The growth
rate and saturation strength of the field are increasing as function
of increasing forcing, and thereby also as functions of the Reynolds
number $\mbox{Rm}$, so that for the highest forcing
investigated, the time scale of growth is roughly 27\,Myrs,
building up the magnetic field in a few hundreds of
Myrs. A similar, but isothermal, system has been investigated earlier by
\cite{Axel01}, who found otherwise similar behavior, but the energy of
the magnetic fields generated in the system exceeded the kinetic
energy of turbulence. The build-up of the large-scale field from the
equipartition to the super-equipartition values was observed to occur
on the diffusive timescale, and therefore be dependent on
$\mbox{Rm}$. Such behavior is not seen in the system of this study.

\begin{figure}[t!]\begin{center}
\includegraphics[width=0.49\textwidth]{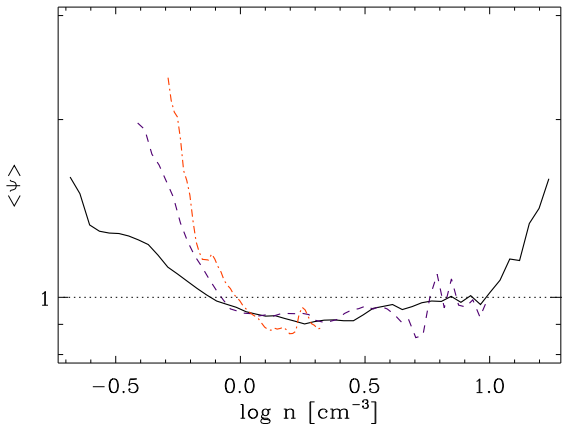}
\end{center}\caption[]{The calculated flux diagnostic
  $\left<\Psi\right>$ from Eq.(\ref{eq:diag}) for the thermally
  stable run TSa (dashed-dotted), TSd (dashed) and TSc (solid line). 
}\label{sflux}\end{figure}

Next, we study the spatial distribution of the magnetic fields, and
convince ourselves that the dynamo seen in the helical forcing case is
indeed of the large-scale type, being capable of generating magnetic
fields on spatial scales larger than the forcing scale of the
turbulence, $k_f/k_1=3$. In the two leftmost panels of
Fig.~\ref{bsnap}, we plot two components of the magnetic field for
Run~TSa after the growth of the magnetic field has saturated. For the
$B_y$ component, a clear sinusoidal wave of the Fourier mode $k=1$
over the $x$-coordinate is seen, while $B_z$ shows the same mode with
a cosinus dependence of the spatial coordinate. The patterns seen are
very similar to the study of \citet{Axel01}, and the field is clearly
coherent over larger scales as the forcing scale. Such a mean field
vanishes when averaging over the whole volume, but the rms-strength of
the large-scale component $\sqrt{\left<B_y^2\right>_{yz}}$ can be
recovered if averaging over $y$- and $z$-directions, denoted with
$\left<\right>_{yz}$, is used; we use averaging of this type in
Table~\ref{table:mgfs} to estimate the generated large-scale field
strengths.

The scales of the large-scale fields are the same for all the runs
independent of forcing, but the direction of the generated mean field
varies almost from one run to another. As can been seen from
Table~\ref{table:mgfs}, the weakest forcing case, Run~TSf, the
Reynolds number being very close to the marginal one and the time
scale of growth very long, exhibits mean fields in every
direction. Run~TSa, the magnetic field shown in Fig.~\ref{bsnap},
develops strong mean components of $B_y$ and $B_z$ in the
$x$-direction. The intermediate forcing cases, Runs~TSd and TSb, on
the other hand, show mean components of $B_x$ and $B_z$ over the
$y$-direction. Finally, the strongest forcing case, Run~TSc, develops
the strongest mean fields in $B_x$ and $B_y$ over the
$z$-direction. Also in the isothermal study of \cite{Axel01}, any
direction of the mean field could be preferred, only depending on the
fine details of the initial random distribution. As is evident when
comparing the rms values of the total magnetic field, presented in
Table~\ref{table:runs}, to the strength of the mean component in
Table~\ref{table:mgfs}, almost all magnetic energy is contained in the
$k=1$ mode. As the forcing is increased, small contribution of the
small-scale field is generated, but in all cases investigated this
contribution is negligible, verifying the assumption that no
small-scale dynamo is present in the models.

When calculating the flux diagnostic, defined by Eq.~(\ref{eq:diag}),
for a magnetic field with a mean component at the wavenumber $k=1$
over a certain direction, having different signs and roughly equal
magnitudes in the different halves of the box, the total magnetic flux
would evidently be close to zero. Therefore, we need to use absolute
values of the field strength, $\left| \bm{B} \right|$, in
  Eq.~\ref{eq:diag}. Furthermore, as the distribution of the mean fields
can be quite complex, we calculate an average value
$\left<\Psi\right>$ of $\Psi^{xy}$, $\Psi^{yz}$, and $\Psi^{xz}$. In
Fig.~\ref{sflux} we show plots of the magnetic flux diagnostic for the
thermally stable runs developing dynamo action. For the weakest
forcing case, Run~TSa, the magnetic and density fields show signs of
anti-correlation for low densities, $n<1$cm$^{-3}$, and no clear
correlation for densities above that limit. The anti-correlation at
lower densities is observed to decrease with increasing forcing and
dynamo efficiency, but only for the highest forcing and most efficient
dynamo, Run~TSc, are signs of positive correlation at higher densities,
10-35\,cm$^{-3}$, established. This clearly shows that the
compression due to the forcing alone is very inefficient in producing
any significant positive correlation between the density and magnetic
field.

\begin{figure}[!ht]\begin{center}
    \includegraphics[width=0.98\columnwidth]{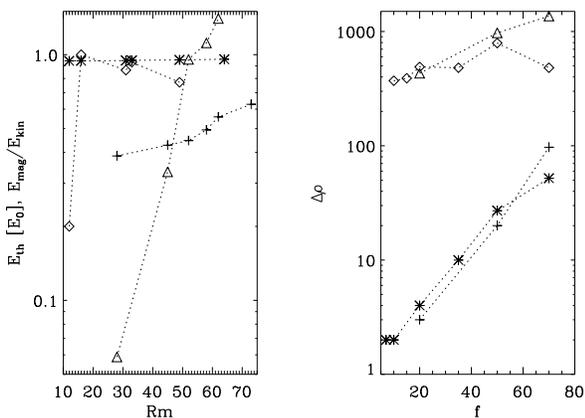}
    \caption[]{Left panel: Thermal energies and the ratio of magnetic
      to kinetic energy as function of magnetic Reynolds number for
      runs with different cooling functions. Stars: thermal energy in
      the thermally stable MHD Runs~TSf -- TSc. Crosses: thermal
      energy in the thermally unstable MHD Runs~SSf -- SSc. Diamonds:
      magnetic to kinetic energy ratio in Runs~TSf--TSc. Triangles:
      magnetic to kinetic energy ratio in Runs~SSf--SSc. Right panel:
      Density contrast $\Delta \rho$ from MHD runs (stable with stars,
      unstable with diamonds) compared with their hydrodynamic
      counterparts (stable with crosses, unstable with
      traingles) as function of the forcing amplitude, $f$.}\label{fig:rm_dep}
\end{center}
\end{figure}

\begin{figure*}[!ht]\begin{center}
\includegraphics[width=0.33\textwidth]{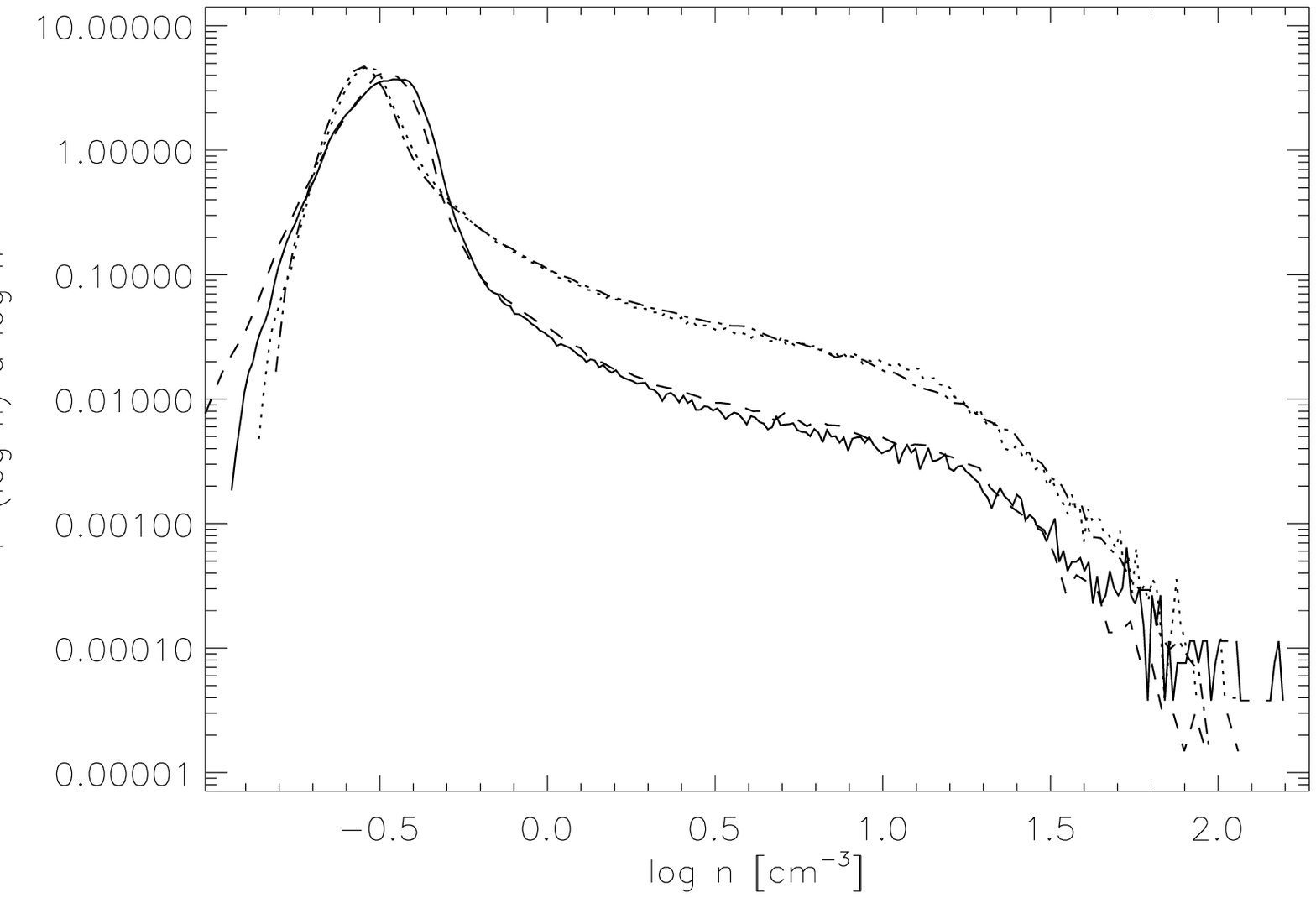}\includegraphics[width=0.33\textwidth]{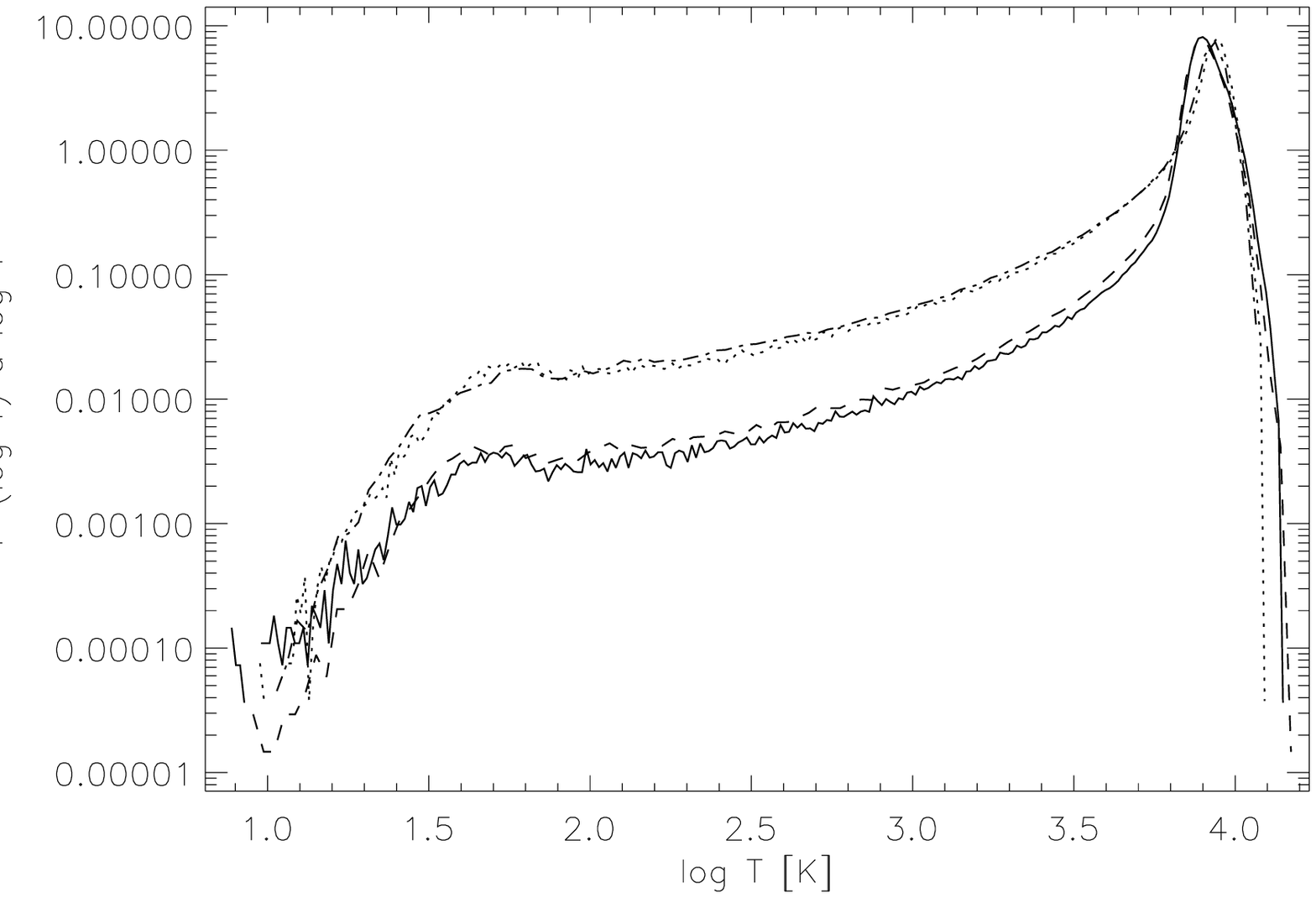}\includegraphics[width=0.33\textwidth]{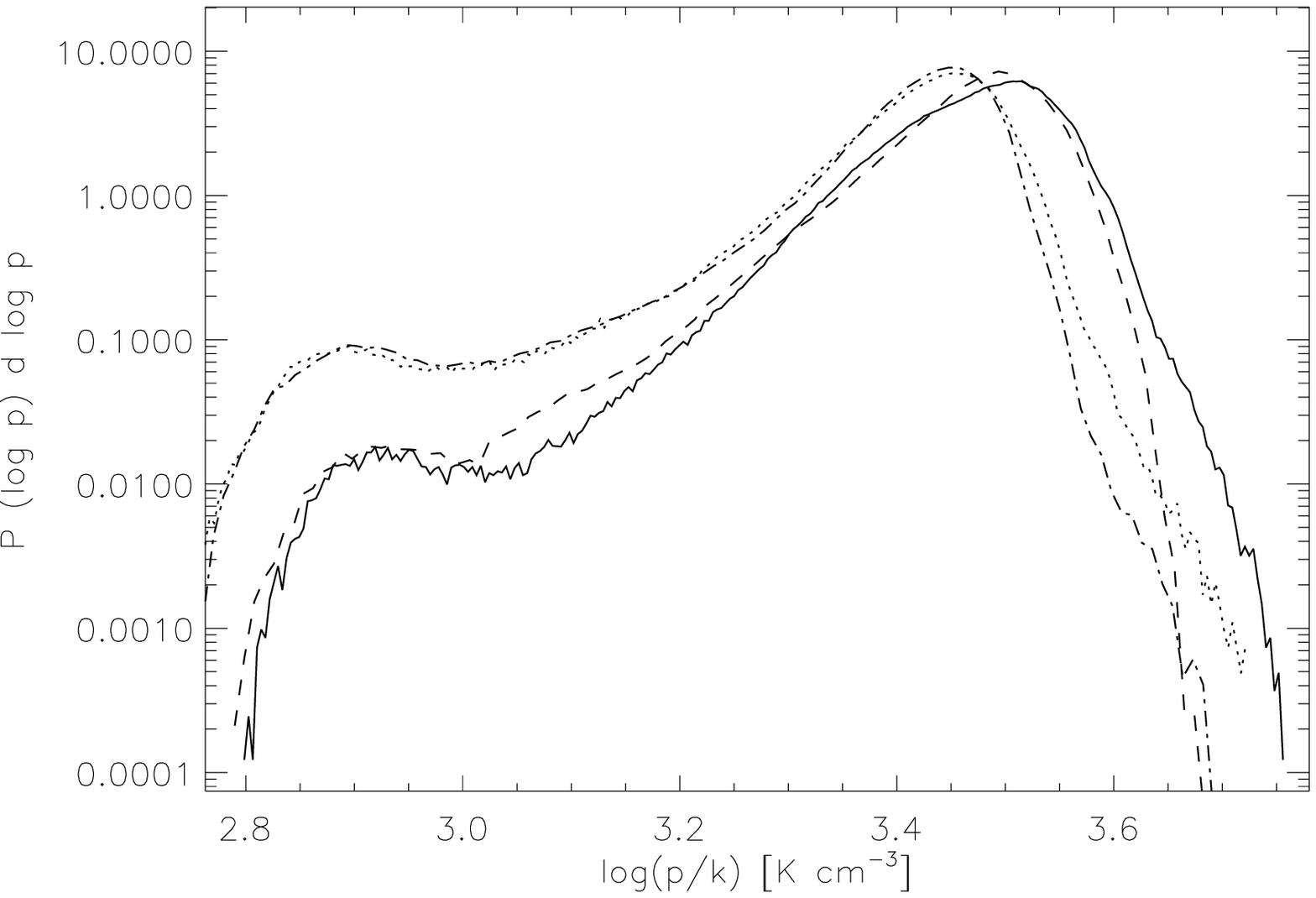}
\end{center}\caption[]{Probability density functions for the thermally
  unstable hydrodynamic runs HDSSc (solid line) compared with
  the magnetohydrodynamic counterpart SSc (dashed line), and
  HDSSb (dotted line) with SSb (dashed-dotted line). On the
  left we show the PDF of logarithmic density, in the middle the PDF
  of logarithmic temperature, and on the right the PDF of
  logarithmic pressure.}\label{mhd_pdfs}\end{figure*}

\subsection{Thermally unstable magnetohydrodynamic runs}

Finally, we produce magnetohydrodynamic runs with helical forcing
using the thermally unstable cooling function, the results being
listed in the lower part of Tables~\ref{table:runs} and
\ref{table:mgfs} and the time evolution of the magnetic energy for
  some runs shown in Fig.~\ref{fig:emag}. As is evident from this table,
when comparing these runs with their thermally stable counterparts with the
same magnitude of forcing, it can be noted that the turbulence is more
vigorous with the thermally unstable cooling function in the sense
that rms values of velocity and Reynolds numbers are larger for any
fixed forcing amplitude. The kinetic energy, however, is smaller for
the thermally unstable runs than for the thermally stable ones,
explained by the density distribution being dominated by the warm
diffuse gas with low density for the thermally unstable setup.

For the Run~SSf with the forcing amplitude $f$=15 producing turbulence
with $\mbox{Rm} \approx 45$, a very slowly growing dynamo is found, as
seen from Fig.~\ref{fig:emag}, whereas the magnetic field of the
Run~SSe with $\mbox{Rm} \approx 28$ is seen to decay
exponentially. The critical Reynolds number, therefore, lies
somewhere in between 28--45, roughly twice the value found for the
thermally stable runs. In other words, the large-scale dynamo is
harder to excite in the thermally unstable system. As is evident from the
left panel of Fig.~\ref{fig:rm_dep}, where we plot the dependence of
thermal energy and the ratio of magnetic to kinetic energy ratio on
the magnetic Reynolds number $\mbox{Rm}$, the magnetic energy in the
saturated state remains below the kinetic energy of turbulence up to
the forcing amplitude $f$=35 in Run~SSd, for which the Reynolds number
is 58. For the thermally stable runs the dynamo efficiency of this
level was reached with $f$=20 with $\mbox{Rm}$=27, the difference
being again by a factor of two. For the {\rm strongest} forcings
investigated (Runs~SSb and SSc), the magnetic energy exceeds the
kinetic energy of turbulence. Such behavior was not seen in the
thermally stable runs.

The density contrast can be seen to become affected by the presence of
the magnetic field; this effect is visible in Fig.~\ref{fig:rm_dep}
right panel, where we plot $\Delta \rho$ both for the MHD runs and
their hydrodynamic counterparts. For both the stable and unstable runs
the trend appears very similar: as long as the magnetic field energy
remains clearly below the kinetic energy of turbulence, i.e. the
dynamo efficiency is still low, the density contrast becomes
enhanced. For forcings capable of generating a magnetic field close to
or exceeding the kinetic energy of turbulence (Runs~TSc, SSb, and SSc)
the density contrast becomes reduced. This can be interpreted as the
lowered capability of either the turbulence or the condensation mode
of the thermal instability to create density structures in the
presence of a strong dynamo-generated magnetic field, in agreement
with the results from earlier magnetohydrodynamic studies without
self-consistent dynamo action, e.g. by \citet{ostriker99} and
\citet{padoan99}.

In Fig.~\ref{mhd_pdfs} we show PDFs of the magnetohydrodynamic runs
with the strongest forcings (Runs~SSb and SSc) compared to the
hydrodynamic counterparts (Runs~HDSSb and HDSSC). The main effect due
to the presence of the magnetic field seen in them is the decreased
amount of dense gas at the very high end of the density distribution,
and a slight increase of very low density gas for Run~SSc, while
hardly any difference can be seen in the temperature distributions. The
pressure PDF, therefore, reflects the decreased densities, the higher
pressure wing being reduced for both Runs~SSb and SSc. This implies
that the stronger the magnetic field grows, the more it resists the
cloud compression due to thermal instability. In a similar study,
\citet{Gazol09} reported that an initially uniform magnetic field of
varying strengths caused an extended low-pressure wing to their
pressure distribution; in our case, the wing was seen already in the
hydrodynamic regime, and the effect of the magnetic field on it is
barely visible. In a related, but not completely comparable, study by
\citet{Avillez05}, the effect of the presence of a magnetic field on a
supernova-regulated flow was studied. In contrast to our results,  the
amount of cold and dense gas was observed to increase in the
magnetohydrodynamic versus the hydrodynamic setup. The differences
seen between our model to the other two might be connected to the fact
that in our case self-consistently dynamo-active flows were studied in
contrast to imposed ones passively advected by the flow.

\begin{figure}[!ht]\begin{center}
\includegraphics[width=0.49\textwidth]{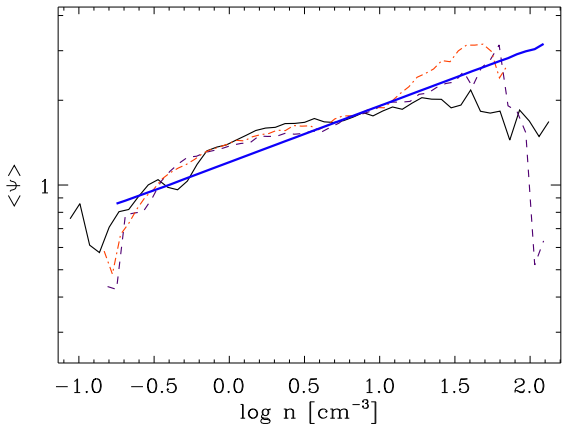}
\end{center}\caption[]{The calculated flux diagnostic
  $\left<\Psi\right>$ from Eq.(\ref{eq:diag}) for the thermally
  unstable runs SSd (dashed-dotted), SSb (dashed) and SSc
  (solid line). The thickest line shows a fit of the type $B
  \propto \rho^{0.2}$. }\label{uflux}\end{figure}

Finally, we plot the flux diagnostic $\left< \Psi \right>$ for the
dynamo-active thermally unstable runs in Fig.~\ref{uflux}. If compared
to the corresponding plot for thermally stable runs, Fig.~\ref{sflux}
showing uncorrelated density and magnetic fields except for the
high-density end of the run with the highest forcing, the action of
thermal instability on the system produces clear positive correlation
between the two quantities. The correlation for all the cases
investigated is considerably weaker, of the form $B \propto
\rho^{0.2}$ (thick line plotted in the figure), compared to the one
observed in the interstellar matter. The strongest correlation is seen
for the Run~SSd, the trend declining somewhat for higher forcings. For
all the runs, there seems to be a tendency for the correlation to
break down at the highest densities.

\section{Conclusions}

In this paper we have investigated the role of thermal instability for
large-scale dynamo action in galaxies and in producing the
density-magnetic field correlation in the interstellar matter. A
periodic cubic magnetohydrodynamic setup with external helical forcing
was used to establish large-scale dynamo action in the system,
which was subject either to thermally stable or unstable cooling
functions having the same integrated net cooling for constant
density. Reference runs without magnetic fields and with non-helical
forcing were made, the former to isolate the effect of the magnetic
fields on the system, and the latter for detecting if small-scale
dynamo action was produced in the Reynolds number regime
investigated. The latter set of runs showed that no small-scale dynamo
action was to be expected in the helical runs, the critical Reynolds
number for small-scale dynamo action in this particular system being
larger than 97. This number is larger than obtained in similar, but
isothermal, setups investigated.

In the thermally stable runs, the critical Reynolds number of
large-scale dynamo action was calculated to be in between 12-16, whereas
for the thermally unstable setup the corresponding range was found to
be roughly twice that, 28-45; the thermal instability, therefore,
makes the onset of the large-scale dynamo harder. In contrast to the
thermally stable cases investigated, that were observed to produce
magnetic fields in equipartition with the kinetic energy of
turbulence, the thermally unstable ones were capable of producing
magnetic fields with somewhat larger energy than contained in the
turbulent velocity field. In terms of the rms-magnetic field strength,
the thermally stable runs produced roughly 1.5 times larger magnetic
fields than the thermally unstable counterparts per fixed forcing, the
maximal values being of the order of 1.5$\mu$G for the strongest
forcing. The dynamos, no matter which cooling function was used,
always produced magnetic fields with a mean component at the
wavenumber $k$=1. The large-scale field could be generated in any
direction, and show systematic variation over any spatial coordinate,
depending on the strength of the forcing.

In the thermally stable runs the denser structures were formed due to
the action of the turbulent velocity field itself. The enhanced
density contrast produced by increasing forcing, however, was found to
be very inefficient in producing a density-magnetic field correlation in
the system. In fact, a positive correlation was found only for the
highest forcing case investigated, and even then only at the limit of
the highest densities. For all the other cases, our flux diagnostic
indicated a tendency for uncorrelation between the quantities. For the
thermally unstable dynamo-active cases investigated, a positive
correlation of the form $B \propto \rho^{0.2}$ was observed to be
established, but a tendency of this rather weak correlation to break
down at the high density limit was seen. It can be concluded that an
instability of any 'condensation' type, i.e. the gravitational or
  condensation mode of thermal instability, is instrumental in
establishing density-magnetic field correlation in the ISM, but based
on this study, the correlation predicted due to thermal instability is
weak, and especially so for high densities.

\acknowledgements The fruitful collaboration with Prof. Anvar Shukurov
and Dr. Andrew Fletcher, from which this work has significantly
benefited, is gratefully acknowledged. Fruitful discussions with
Doc. Jorma Harju, Prof. Axel Brandenburg and Doc. Petri K\"apyl\"a
have also greatly helped in preparing this manuscript. We also
acknowledge the anonymous referee, whose insightful comments helped us
to significantly improve the manuscript. This work was supported by
the Academy of Finland through the grants No.\ 112020, 141017, and
218159. All the computations have been performed in the supercomputers
hosted by the Center for Scientific Computing Ltd. (CSC), which
organization is gratefully acknowledged for the granted CPU time.
\bibliographystyle{apj}      
\bibliography{apj-jour,mara}

\end{document}